\documentclass[12pt]{article}
\usepackage{amsmath,amsthm}

\begin{document}
\vspace*{3cm}
\begin{center}
{\LARGE{\bf 
Superspace formulation of general massive\\ 

\smallskip 
gauge theories and geometric interpretation\\

\medskip
of mass--dependent BRST symmetries }}
\\
\bigskip\bigskip\bigskip
{\large{\sc B. Geyer}}\footnote{e-mail: geyer@itp.uni-leipzig.de}
\\
\smallskip
{\it Universit\"at Leipzig, Naturwissenschaftlich--Theoretisches Zentrum
and \\Institut f\"ur Theoretische Physik,
 04109 Leipzig, Germany}
\\
\bigskip
{\large{\sc D. M\"ulsch}}
\\
\smallskip
{\it Wissenschaftszentrum Leipzig e.V., Leipzig 04103, Germany}
\\
\bigskip\bigskip\bigskip\bigskip
{
{\bf Abstract}}
\medskip
\\
\end{center}

\begin{quotation}
\noindent {\small{A superspace formulation is proposed for the 
$osp(1,2)$--covariant Lagrangian quantization of general massive gauge 
theories. The superalgebra $osp(1,2)$ is considered as subalgebra of 
the superalgebra $sl(1,2) \sim osp(2,2)$ which may be considered as the 
algebra of generators of the conformal group in a superspace with two 
anticommuting coordinates. The mass--dependent (anti)BRST symmetries of 
proper solutions 
of the quantum master equations in the osp(1,2)--covariant formalism are 
realized in that superspace as invariance under translations combined with 
mass--dependent special conformal transformations. The $Sp(2)$ symmetry --
in particular the ghost number conservation -- and the new ghost number 
conservation are realized in the superspace as invariance under symplectic 
rotations and dilatations, respectively. The new ghost number conservation 
is generally broken by the choice of a gauge. The transformations of the 
gauge fields and the full set of necessarily required (anti)ghost and 
auxiliary fields under the superalgebra $sl(1,2)$ are determined both for 
irreducible and first--stage reducible theories with closed gauge algebra.}}
\end{quotation}

\bigskip
\bigskip
\pagebreak
\setlength{\baselineskip}{0.7cm}


\section{Introduction}
\setcounter{equation}{0}


After the realization that the effective Lagrangian of non--abelian
gauge theories is invariant with respect to Becchi-Rouet-Stora--Tyutin
(BRST) \cite{1} as well as  
anti--BRST transformations \cite{1a}, it has been recognized that
this invariance can be used as a fundamental principle in the construction
of covariantly quantized gauge theories (for a modern introduction see
\cite{1b}). In particular, a superfield 
formulation of quantized pure Yang--Mills theories by Bonora and Tonin 
provides a convenient framework for describing the extended BRST symmetries 
\cite{2}. In this framework the extended BRST symmetries are realized as 
translations in a superspace along additional anticommuting coordinates 
(for a more recent approach, we refer to \cite{3} and references therein).

A $Sp(2)$--covariant superfield description of Lagrangian quantization
of general gauge theories, which is applicable irrespective of 
whether the theories are irreducible or reducible and whether 
the gauge algebra is closed or open, has been given in Ref. \cite{4}. 
A corresponding superfield formulation of the quantization procedure 
in the Hamiltonian approach for theories with first--class constraints 
has been given in Ref.~\cite{5}.

Recently, the $Sp(2)$--quantization of Batalin, Lavrov and Tyutin(BLT)
has been extended to a formalism which is based on the orthosymplectic 
superalgebra $osp(1,2)$ \cite{6} and which can be applied to {\it massive} 
gauge theories. This is achieved by incorporating into the extended BRST 
transformations $m$--dependent terms in such a way that the $m$--extended 
(anti)BRST symmetry of the quantum action $W_m$ is preserved. 
In that approach $W_m$ is required to satisfy the
generalized quantum master equations of $m$--extended BRST symmetry 
and, in addition, of $Sp(2)$ symmetry,
\begin{eqnarray}
\label{qme1}
\hbox{$\frac{1}{2}$} ( W_m, W_m )^a + V_m^a W_m 
&=& i \hbar \Delta^a W_m
\quad\Longleftrightarrow\quad
\bar{\Delta}^a \exp\{W_m\}=0,
\\ 
\label{qme2}
\hbox{$\frac{1}{2}$} \{ W_m, W_m \}_\alpha + V_\alpha W_m 
&=& i \hbar \Delta_\alpha W_m
\quad\Longleftrightarrow\quad
\bar{\Delta}_\alpha \exp\{W_m\}=0,
\end{eqnarray}
respectively, whose generating (second order) differential operators
\begin{eqnarray} 
\label{Delta1}
\bar{\Delta}_m^a &\equiv& \Delta^a + (i/\hbar) V_m^a,
\qquad (a = 1,2),\\ 
\label{Delta2}
\bar{\Delta}_\alpha &\equiv& \Delta_\alpha + (i/\hbar) V_\alpha,
\qquad 
(\alpha = 0, \pm 1) ,
\end{eqnarray}
form a superalgebra isomorphic to $osp(1,2)$ (the definitions of the
(anti)brackets and the operators $\bar{\Delta}_m^a$ and 
$\bar{\Delta}_\alpha$ are given below). 

The incorporation of mass terms into the action $W_m$ is necessary at least 
intermediately in the renormalization scheme of Bogoliubov, Parasiuk, Hepp,
Zimmermann and Lowenstein(BPHZL)\cite{BPHZL}. 
An essential ingredient to deal with 
massless theories in that scheme consists in the introduction of a 
regularizing mass $m = (s - 1) M$ for any massless field and performing
ultraviolet as well as infrared subtractions thereby avoiding
spurious infrared singularities in the limit $s \rightarrow 1$. By using 
such an infrared regularization -- without violating the extended BRST 
symmetries -- the $osp(1,2)$--superalgebra appears necessarily.

Moreover, the BPHZL renormalization scheme is probably the mathematical best
founded one in order to formulate the quantum master equations on the level
of algebraic renormalization theory and to properly compute
higher--loop anomalies \cite{7}. The reason is the following:
The only quantity that remains undefined 
in the above mentioned approaches of quantizing general gauge theories
is the right--hand side of the quantum master equations (that problem already
occurs in the Batalin--Vilkovisky(BV) field--antifield formalism). At the 
classical level, the extended BRST invariance in the $osp(1,2)$--approach 
is expressed by the classical master equations 
$\hbox{$\frac{1}{2}$} ( S_m, S_m )^a + V_m^a S_m = 0$, where $S_m$ is the
lowest order approximation in $\hbar$ of $W_m$. On the quantum level, 
formal manipulations modify the classical master equations into 
Eq.~(\ref{qme1}).
When applied to the local functional $W_m$ 
the operation $\Delta^a W_m$ leads to the ill--defined expression $\delta(0)$.
Well--defined expressions for the regularized operators $\Delta^a$ are 
proposed at one--loop level in \cite{8} within the context of Pauli-Villars
regularization and at higher order in \cite{9} for non--local regularization.
However, by means of the BPHZL renormalization scheme, which bypasses any 
ultraviolet regularizations, the right--hand side of the quantum master 
equations can be defined by using Zimmermanns's normal products to any 
order of perturbation theory \cite{7}.

The purpose of the present paper is to reveal the geometrical content of the
$osp(1,2)$--covariant Lagrangian quantization which amounts to understand
the geometrical meaning of the $m$--dependent part of the extended BRST
transformations. For that reason the theory will be described in terms of
super(anti)fields. Our approach is based on the idea to consider 
$osp(1,2)$ as subsuperalgebra of the superalgebra $sl(1,2)$. The latter 
algebra, being isomorphic to $osp(2,2)$, contains four bosonic generators 
$V_\alpha$ and $V$, which form the Lie algebra $sl(2) \oplus u(1)$, and 
four (nilpotent) fermionic generators $V_+^a$ and $V_-^a$. 
The even part of $osp(1,2)$ is the algebra $sl(2)$  
generating the special linear transformations, but due to their isomorphism
to the algebra $sp(2)$ we will speak about symplectic transformations. 
The eigenvalues of the generators $V_\alpha$ for $\alpha = 0$ define 
the ghost numbers, whereas the eigenvalues of the
generator $V$ define what in Ref.~\cite{10} was called the `new ghost
number'. The generators $V_+^a$ and $V_-^a$ have opposite new ghost numbers,
${\rm ngh}(V_\pm^a) = \pm 1$, respectively. But, introducing a mass 
$m$ which formally will be attributed also by a new ghost number,
${\rm ngh}(m) = 1$, they can be combined into two fermionic 
generators $V_m^a = V_+^a + \hbox{$\frac{1}{2}$} m^2 V_-^a$ of the 
superalgebra $osp(1,2)$. For $m\neq 0$ these generators $V_m^a$ 
are neither nilpotent nor do they anticommutate among themselves.

The key observation that allows for a geometric interpretation of the
superalgebra $sl(1,2)$ is due to Baulieu, Siegel and Zwiebach \cite{11}
which in a
quite different context of string theory gave a description of $sl(1,2)$
as the algebra generating conformal transformations in a 2--dimensional
superspace. Hence, the generators $V_+^a, V_-^a,
V^{ab} =  (\sigma^\alpha)^{ab} V_\alpha$ and $V$ of the superalgebra 
$sl(1,2)$, with $(\sigma^\alpha)^{ab}$ generating the fundamental
representation of $sl(2)$, may be considered as generators of translations 
$i P^a$, special conformal transformations $i K^a$, symplectic rotations 
$i M^{ab}$ and dilatations $-i D$, respectively, in superspace. 
This leads immediatly to a `natural' geometric formulation of the $osp(1,2)$ 
quantization procedure: 
In a superspace description the invariance of $W_m$ under $m$--extended 
BRST transformations, generated by 
$V_m^a = V_+^a + \hbox{$\frac{1}{2}$} m^2 V_-^a$, corresponds to translations 
combined with $m$--dependent special conformal transformations, and
 its invariance 
under $Sp(2)$--transformations, generated by $V_\alpha$, corresponds to 
symplectic rotations. Furthermore, solutions $S_m$ of the classical master 
equations $\hbox{$\frac{1}{2}$} ( S_m, S_m )^a + V_m^a S_m = 0$ and 
$\{ S_m, S_m \}_\alpha + V_\alpha S_m = 0$ with vanishing new ghost number,
${\rm ngh}(S_m) = 0$, correspond to solutions in the superspace being 
invariant under dilatations, generated by $V$.

The paper is organized as follows. In Sect.~2 we shortly review some basic
definitions and properties of $L$--stage reducible gauge theories and we 
introduce the corresponding configuration space of fields and antifields. 
Furthermore, the (anti)commutation relations of the superalgebra $sl(1,2)$ 
are defined and an explicit realization in terms of linear differential
operators acting on the antifields are given. 
In Sect.~3 the superalgebra $sl(1,2)$ is realized as algebra of the 
conformal group in superspace where the usual space--time is extended
by two extra anticommuting coordinates $\theta^a$. Moreover, we give 
a superspace representation of the algebra $sl(1,2)$ acting linearly 
on the super(anti)fields. 
In Sect.~4 the $osp(1,2)$--covariant superfield quantization rules for 
general gauge theories are formulated. Besides, it is shown that proper
solutions of the classical master equations can be constructed being 
invariant under $osp(1,2) \oplus u(1)$, where the additional $u(1)$ symmetry
is related to the new ghost number conservation; however, this symmetry is
broken by choosing a gauge. Sect.~5 is devoted to study the (in)dependence 
of general Green's functions on the choice of the gauge. In the $osp(1,2)$
approach it is proven that mass terms generally destroy gauge independence; 
however, this gauge dependence disappears in the limit $m = 0$. 
In Sect.~6 we construct $osp(1,2) \oplus u(1)$ symmetric proper solutions
of the classical master equations. Moreover, the problem of how to determine
the transformations of the gauge fields and the full set of the necessary
(anti)ghost and auxiliary fields under the superalgebra $sl(1,2)$ is 
solved both for irreducible and first--stage reducible theories with closed
algebra.

Throughout this paper we have used the condensed notation introduced by
DeWitt \cite{12} and conventions adopted in Ref.~\cite{6}; if not specified 
otherwise, derivatives with respect to the superantifields 
$\bar{\Phi}_A(\theta)$ and the superspace coordinates $\theta^a$ are the 
(usual) left ones and that with respect to the superfields $\Phi^A(\theta)$ 
are {\it right} ones. Left derivatives with respect to $\Phi^A(\theta)$ and 
right derivatives with respect to $\theta^a$ are 
labelled by the subscript $L$ and $R$, respectively; for example, 
$\delta_L/ \delta \Phi^A(\theta)$ ($\partial_R/ \partial \theta^a$) 
denotes the left(right) derivative with respect to the superfields 
$\Phi^A(\theta)$ (the superspace coordinates $\theta^a$).  


\section{Realization of $sl(1,2)$ in terms of antifields}
\setcounter{equation}{0}


\noindent (A) {\it General gauge theories }
\\
Before going into the main subject of this section let us shortly introduce the 
basic definitions of general gauge theories and the corresponding
configuration space of fields and antifields:
 
A set of gauge (as well as matter) fields $A^i$ with Grassmann parities 
$\epsilon(A^i) = \epsilon_i$ will be considered whose classical action 
$S_{\rm cl}(A)$ is invariant under the gauge transformations 
\begin{equation}
\label{II1}
\delta A^i = R^i_{\alpha_0} \xi^{\alpha_0},
\qquad
\alpha_0 = 1, \ldots, n_0,
\qquad
S_{{\rm cl}, i} R^i_{\alpha_0} = 0;
\end{equation}
here, $\xi^{\alpha_0}$ are the parameters of these transformations and 
$R^i_{\alpha_0}(A)$ are the gauge generators having Grassmann parity 
$\epsilon(\xi^{\alpha_0}) = \epsilon_{\alpha_0}$ and  
$\epsilon(R^i_{\alpha_0}) = \epsilon_i + \epsilon_{\alpha_0}$, respectively;
by definition $X_{, j} = \delta X/ \delta A^j$.

For {\it general gauge theories} the algebra of generators has the form 
\cite{10}:
\begin{equation}
\label{II2}
R_{\alpha_0, j}^i R_{\beta_0}^j -
(-1)^{\epsilon_{\alpha_0} \epsilon_{\beta_0}} 
R_{\beta_0, j}^i R_{\alpha_0}^j =
- R_{\gamma_0}^i F_{\alpha_0 \beta_0}^{\gamma_0} - 
M^{ij}_{\alpha_0 \beta_0} S_{{\rm cl}, j},
\end{equation}
where $F_{\alpha_0 \beta_0}^{\gamma_0}(A)$ are the field--dependent 
structure functions and the matrix $M^{ij}_{\alpha_0 \beta_0}(A)$ is graded 
antisymmetric with respect to $(ij)$ and $(\alpha_0 \beta_0)$. The gauge 
algebra is said to be {\it closed} if $M_{\alpha_0 \beta_0}^{ij} = 0$,
otherwise it is called {\it open}. Moreover, Eq. (\ref{II2}) defines a 
Lie algebra if the algebra is closed and the 
$F_{\alpha_0 \beta_0}^{\gamma_0}$ do not depend on $A^i$.

If the set of generators $R^i_{\alpha_0}$ are linearly {\it independent} then 
the theory is called {\it irreducible} \cite{13}. On the other hand, 
if the generators $R^i_{\alpha_0}$ are not independent, i.e., 
if on--shell certain relations exist among them, then,
according to the following characterization, the theory under consideration 
is called $L$--stage {\it reducible} \cite{14}: \\
There exists a chain of field--dependent on--shell zero--modes 
$Z^{\alpha_s - 1}_{\alpha_s}(A)$,
\begin{alignat*}{2}
R^i_{\alpha_0} Z^{\alpha_0}_{\alpha_1} &= 
S_{{\rm cl}, j} K^{ji}_{\alpha_1},
&\qquad
K^{ij}_{\alpha_1} &= - (-1)^{\epsilon_i \epsilon_j} K^{ji}_{\alpha_1},
\\
Z^{\alpha_{s - 2}}_{\alpha_{s - 1}} Z^{\alpha_{s - 1}}_{\alpha_s} &= 
S_{{\rm cl}, j} K^{j \alpha_{s - 2}}_{\alpha_s},
&\qquad
\alpha_s &= 1, \ldots, n_s, ~ s = 2, \ldots, L,
\end{alignat*}
where the stage $L$ of reducibility is defined by the lowest value $s$ 
for which the matrix $Z^{\alpha_{L - 1}}_{\alpha_L}(A)$ is no longer 
degenerated. The $Z^{\alpha_{s - 1}}_{\alpha_s}$ are the on--shell zero 
modes for $Z^{\alpha_{s - 2}}_{\alpha_{s - 1}}$ with 
$\epsilon(Z^{\alpha_{s - 1}}_{\alpha_s}) = \epsilon_{\alpha_{s - 1}} + 
\epsilon_{\alpha_s}$.
In the following, if not stated otherwise, we assume $s$ to
take on the values $s = 0, \ldots, L$, thereby including also the case of
irreducible theories. 

The whole space of fields $\phi^A$ and antifields $\bar\phi_A, 
\phi^*_{Aa}, \eta_A$ together with their Grassmann parities 
(modulo 2) is characterized by the following sets \cite{10,6}
\begin{alignat*}{2}
\phi^A &= ( A^i, B^{\alpha_s| a_1 \cdots a_s},  
C^{\alpha_s| a_0 \cdots a_s}, s = 0, \ldots L ), 
&\qquad
\epsilon(\phi^A) &\equiv \epsilon_A = 
( \epsilon_i, \epsilon_{\alpha_s} + s, \epsilon_{\alpha_s} + s + 1 )
\\
\bar{\phi}_A &= ( \bar{A}_i, \bar{B}_{\alpha_s| a_1 \cdots a_s}, 
\bar{C}_{\alpha_s| a_0 \cdots a_s}, s = 0,\ldots L ),
&\qquad
\epsilon(\bar{\phi}_A) &= \epsilon_A,
\\
\phi^*_{A a} &= ( A^*_{i a}, B^*_{\alpha_s a| a_1 \cdots a_s},
C^*_{\alpha_s a| a_0 \cdots a_s} , s = 0, \ldots L ),
&\qquad
\epsilon(\phi^*_{A a}) &= \epsilon_A + 1,
\\
\eta_A &= (D_i, E_{\alpha_s| a_1 \cdots a_s}, 
F_{\alpha_s| a_0 \cdots a_s}, s = 0,\ldots L),
&\qquad
\epsilon(\eta_A) &= \epsilon_A,
\end{alignat*}
respectively.
Here, the pyramids of auxiliary fields $B^{\alpha_s| a_1 \cdots a_s}$ and  
(anti)ghosts $C^{\alpha_s| a_0 \cdots a_s}$  are 
$Sp(2)$--tensors of rank $s$ and $s + 1$, respectively, 
being completely {\it symmetric} with 
respect to the `internal' $Sp(2)$--indices $a_i=1,2,\;(i=0,1,\ldots,s)$;
similarly for the
antifields $\bar\phi_A, \phi^*_{Aa}$ and sources $\eta_A$. 
The independent index $a=1,2$
which counts the two components of a $Sp(2)$--spinor will be called 
`external'.
The totally symmetrized $Sp(2)$--tensors are irreducible and have maximal 
$Sp(2)$--spin.
Raising and lowering of $Sp(2)$--indices is obtained by the invariant tensor 
\begin{equation*}
\epsilon^{ab} = \begin{pmatrix} 0 & 1 \\ -1 & 0 \end{pmatrix},
\qquad
\epsilon^{ac} \epsilon_{cb} = \delta^a_b.
\end{equation*}

\smallskip
\noindent (B) {\it The superalgebra sl(1,2) }
\\ 
The main goal of this Section is to determine the action of the 
generators of the superalgebra 
$sl(1,2)$ on the antifields $\bar{\phi}_A$, $\phi_{A a}^*$ and $\eta_A$. 
Let us now introduce that algebra.

The even part of $sl(1,2) \sim sl(2,1)$ is the Lie algebra 
$sl(2) \oplus u(1)$. We denote by $V_\alpha$, ($\alpha = 0, \pm$) the (real)
generators of $SL(2)$ and by $V$ the generator of $U(1)$. The odd part of
$sl(1,2)$ contains two (nilpotent) $SL(2)$--spinors, $V_\pm^a$, with spin
$\frac{1}{2}$ and Weyl weight $\alpha(V_\pm^a) = \pm 1$, respectively.
Spin and Weyl weight of $V_\pm^a$ are defined through their behaviour
under the action of the generators $V_\alpha$ and $V$, respectively.
\footnote{
Identifying $V\equiv iD$ with $D$ being the generator of dilatations
in superspace, as will be done in Sect.~3, the Weyl weight coincides 
with the superspace scale dimension of the corresponding quantity.
Of course, the latter should not be confuced with the scale dimension
of any quantity in ordinary space--time.} 

The (anti)commutation relations of the superalgebra $sl(1,2)$ are 
\cite{15}: 
\begin{alignat}{3}
\label{II3}
[ V, V_\alpha ] &= 0,
&\qquad
[ V, V_+^a ] &= V_+^a,
&\qquad
[ V, V_-^a ] &= - V_-^a,
\nonumber
\\
[ V_\alpha, V_\beta ] &= \epsilon_{\alpha\beta}^{~~~\!\gamma} V_\gamma, 
&\qquad 
[ V_\alpha, V_+^a ] &= V_+^b (\sigma_\alpha)_b^{~a}, 
&\qquad 
[ V_\alpha, V_-^a ] &= V_-^b (\sigma_\alpha)_b^{~a}, 
\\ 
\{ V_+^a, V_+^b \} &= 0,
&\qquad
\{ V_-^a, V_-^b \} &= 0,
&\qquad
\{ V_+^a, V_-^b \} &= - (\sigma^\alpha)^{ab} V_\alpha - \epsilon^{ab} V, 
\nonumber
\end{alignat}
where the $Sp(2)$--indices are raised or lowered according to
\begin{gather*}
(\sigma_\alpha)^{ab} = \epsilon^{ac} (\sigma_\alpha)_c^{~b} =
(\sigma_\alpha)^a_{~c} \epsilon^{cb} =
\epsilon^{ac} (\sigma_\alpha)_{cd} \epsilon^{db},
\\
(\sigma_\alpha)_a^{~b} = - (\sigma_\alpha)^b_{~a},
\qquad~
(\sigma_\alpha)^{ab} = (\sigma_\alpha)^{ba}.
\end{gather*}
The matrices $\sigma_\alpha (\alpha = 0, \pm)$ generate the (real) Lie algebra
$sl(2)$ being isomorphic to $sp(2)$:
\begin{gather}
\label{II4}
(\sigma_\alpha)_a^{~c} (\sigma_\beta)_c^{~b} = g_{\alpha\beta} \delta_a^b + 
\hbox{$\frac{1}{2}$} \epsilon_{\alpha\beta\gamma} (\sigma^\gamma)_a^{~b},
\qquad
(\sigma^\alpha)_a^{~b} = g^{\alpha\beta} (\sigma_\beta)_a^{~b},
\\
g^{\alpha\beta} = \begin{pmatrix} 1 & 0 & 0 \\  0 & 0 & 2 \\ 0 & 2 & 0 
\end{pmatrix},
\qquad
g^{\alpha\gamma} g_{\gamma\beta} = \delta^\alpha_\beta,
\nonumber
\end{gather}
where $\epsilon_{\alpha\beta\gamma}$ is the totally antisymmetric tensor, 
$\epsilon_{0+-} = 1$. For the generators $\sigma_\alpha$ we may choose 
the representation $(\sigma_0)_a^{~b} = \tau_3$ and 
$(\sigma_\pm)_a^{~b} = - \hbox{$\frac{1}{2}$} (\tau_1 \pm i \tau_2)$, with 
$\tau_\alpha$ ($\alpha = 1,2,3$) being the Pauli matrices. 

Let us now rewrite the $sl(1,2)$--algebra in two equivalent forms, 
both of which being of physical relevance in the following.
First, introducing another basis $V^{ab}$ of the $SL(2)$--generators, namely 
\begin{equation}
\label{II5}
V^{ab} = (\sigma^\alpha)^{ab} V_\alpha,
\end{equation}
and making use of the equalities
\begin{equation*}
(\sigma^\alpha)^{ab} (\sigma_\alpha)_d^{~c}  = 
- \epsilon^{c \{a } \delta^{b\} }_d,
\qquad
\epsilon_{\alpha\beta}^{~~~\!\gamma} 
(\sigma^\alpha)^{ab} (\sigma^\beta)^{cd} =
- \epsilon^{\{c \{a } (\sigma^\gamma)^{ b\} d\} },
\end{equation*}
where the curly brackets $\{ ~ \}$ indicate symmetrization of indices, 
the (anti)commutation relations of $sl(1,2)$ read 
\begin{alignat}{3}
[ V, V^{ab} ] &= 0,
&\qquad
[ V, V_+^a ] &= V_+^a,
&\qquad
[ V, V_-^a ] &= - V_-^a,
\nonumber
\\
\label{II6}
[ V^{ab}, V^{cd} ] &= - \epsilon^{\{c \{a } V^{ b\} d\} }, 
&\qquad 
[ V^{ab}, V_+^c ] &= - \epsilon^{c \{a } V_+^{ b\} }, 
&\qquad 
[ V^{ab}, V_-^c ] &= - \epsilon^{c \{a } V_-^{ b\} }, 
\\ 
\{ V_+^a, V_+^b \} &= 0,
&\qquad
\{ V_-^a, V_-^b \} &= 0,
&\qquad
\{ V_+^a, V_-^b \} &= - V^{ab} - \epsilon^{ab} V. 
\nonumber
\end{alignat}
In that form the superalgebra $sl(1,2)$ may be given a geometric 
interpretation as the algebra of the conformal group in a 2--dimensional 
superspace having two anticommuting coordinates (see Sect.~3 below).  

Secondly, we remark that within the field--antifield formalism not the entire 
$sl(1,2)$--superalgebra will be of physical relevance,
since not any of their generators define symmetry operations of the
quantum action -- only some combinations of them forming a 
orthosymplectic superalgebra $osp(1,2)$ generate symmetries 
(see Sect.~4 below). Therefore, with respect to this let us notice 
the isomorphism between $sl(1,2)$ and $osp(2,2)$ by introducing 
the following two combinations of $V_+^a$ and $V_-^a$,   
\begin{equation*}
O_+^a \equiv V_+^a + \hbox{$\frac{1}{2}$} V_-^a,
\qquad
O_-^a \equiv V_+^a - \hbox{$\frac{1}{2}$} V_-^a.
\end{equation*}
Then for the (anti)commutation relations of the superalgebra $osp(2,2)$ 
we obtain
\begin{alignat*}{3}
[ V, V_\alpha ] &= 0,
&\qquad
[ V, O_+^a ] &= O_-^a,
&\qquad
[ V, O_-^a ] &= O_+^a,
\\
[ V_\alpha, V_\beta ] &= \epsilon_{\alpha\beta}^{~~~\!\gamma} V_\gamma, 
&\qquad 
[ V_\alpha, O_+^a ] &= O_+^b (\sigma_\alpha)_b^{~a}, 
&\qquad 
[ V_\alpha, O_-^a ] &= O_-^b (\sigma_\alpha)_b^{~a}, 
\\ 
\{ O_+^a, O_+^b \} &= - (\sigma^\alpha)^{ab} V_\alpha,
&\qquad
\{ O_-^a, O_-^b \} &= (\sigma^\alpha)^{ab} V_\alpha,
&\qquad
\{ O_+^a, O_-^b \} &= - \epsilon^{ab} V. 
\end{alignat*}
Here, $(V_\alpha, O_+^a)$ as well as $(V_\alpha, O_-^a)$ 
obey two different 
$osp(1,2)$--superalgebras with $(V, O_-^a)$ as well as $(V, O_+^a)$ 
forming an 
irreducible tensor of these algebras, respectively, either of them 
transforming according 
to the same representation. Notice, that both $O_+^a$ and $O_-^a$ are 
neither nilpotent nor do they anticommute among themselves. 

\smallskip
\noindent (C) {\it Representation of sl(1,2) on the antifields}
\\
Now, let us give an explicit {\em linear} realization of the generators 
of the superalgebra (\ref{II3}) by their action on the antifields 
$\bar{\phi}_A$, $\phi_{A a}^*$ and the sources $\eta_A$ 
(a nonlinear realization on the fields $\phi^A$ will be given in Sect.~4). 
\begin{alignat}{2}
V_+^a \bar{\phi}_A &= \epsilon^{ab} \phi^*_{A b},
&\qquad
V_-^a \bar{\phi}_A &= 0,
\nonumber
\\
\label{II7}
V_+^a \phi^*_{A b} &= - \delta^a_b \eta_A,
&\qquad
V_-^a \phi^*_{A b} &= \bar{\phi}_B \bigr(
(\sigma^\alpha)^a_{~b} (\sigma_\alpha)^B_{~~\!A} - 
\delta^a_b {\bar\gamma}^B_A \bigr),
\\
V_+^a \eta_A &= 0,
&\qquad
V_-^a \eta_A &= \phi^*_{B b} \bigr(
(\sigma^\alpha)^{ab} (\sigma_\alpha)^B_{~~\!A} - 
\epsilon^{ab} ({\bar\gamma}^B_A + 2 \delta^B_A) \bigr),
\nonumber
\\
\nonumber
\\
V_\alpha \bar{\phi}_A &= \bar{\phi}_B (\sigma_\alpha)^B_{~~\!A},
&\qquad
V \bar{\phi}_A &= \bar{\phi}_B {\bar\gamma}^B_A,
\nonumber
\\
\label{II8}
V_\alpha \phi_{A b}^* &= \phi_{B b}^* (\sigma_\alpha)^B_{~~\!A} +
\phi_{A a}^* (\sigma_\alpha)^a_{~b},
&\qquad
V \phi^*_{A b} &= \phi^*_{B b} ({\bar\gamma}^B_A + \delta^B_A),
\\
V_\alpha \eta_A &= \eta_B (\sigma_\alpha)^B_{~~\!A},
&\qquad
V \eta_A &= \eta_B ({\bar\gamma}^B_A + 2 \delta^B_A)
\nonumber
\end{alignat}
(for a componentwise notation see Appendix A). In Eqs.~(\ref{II7}), 
(\ref{II8})
we introduced two kinds of matrices which deserve some explanation.
The matrices $(\sigma_\alpha)^B_{~~\!A}$ are generalized $\sigma$--matrices 
acting only on internal $Sp(2)$--indices of the (anti)fields, for example,
\begin{equation*}
\bar{\phi}_B (\sigma_\alpha)^B_{~~\!A} = \Bigr( 0,
\sum_{r = 1}^s \bar{B}_{\alpha_s| a_1 \cdots a_{r - 1} a a_{r + 1} \cdots a_s}
(\sigma_\alpha)^a_{~a_r},  
\sum_{r = 0}^s \bar{C}_{\alpha_s| a_0 \cdots a_{r - 1} a a_{r + 1} \cdots a_s}
(\sigma_\alpha)^a_{~a_r} \Bigr);
\end{equation*}
their general definition is given by 
\begin{equation}
\label{II9}
(\sigma_\alpha)^B_{~~\!A} \equiv \begin{cases} 
\delta^{\beta_s}_{\alpha_s} (s + 1) (\sigma_\alpha)^b_{~a}
S^{b_1 \cdots b_s a}_{a_1 \cdots a_s b}  
& \text{for $A = \alpha_s|a_1 \cdots a_s, B = \beta_s|b_1 \cdots b_s$},
\\
\delta^{\beta_s}_{\alpha_s} (s + 2) (\sigma_\alpha)^b_{~a}
S^{b_0 \cdots b_s a}_{a_0 \cdots a_s b}  
& \text{for $A = \alpha_s|a_0 \cdots a_s, B = \beta_s|b_0 \cdots b_s$},
\\
0 & \text{otherwise},
\end{cases}
\end{equation}
where the symmetrizer $S^{b_0 \cdots b_s a}_{a_0 \cdots a_s b}$ is defined as
\begin{equation*}
S^{b_0 \cdots b_s a}_{a_0 \cdots a_s b} \equiv 
\frac{1}{(s + 2)!} \frac{\partial}{\partial X^{a_0}} \cdots
\frac{\partial}{\partial X^{a_s}} \frac{\partial}{\partial X^b} 
X^a X^{b_s} \cdots X^{b_0},
\end{equation*}
$X^a$ being independent bosonic variables.
These operators, obeying $ S^{b_0 \cdots b_s a}_{c_0 \cdots c_s d} 
S^{c_0 \cdots c_s d}_{a_0 \cdots a_s b} =
S^{b_0 \cdots b_s a}_{a_0 \cdots a_s b} $, possess the additional properties
\begin{align*}
S^{b_0 \cdots b_s a}_{a_0 \cdots a_s b} &= \frac{1}{s + 2} \Bigr(
\sum_{r = 0}^s \delta^{b_r}_{a_0} 
S^{b_0 \cdots b_{r - 1} b_{r + 1} \cdots b_s a}_{a_1 \cdots a_s b} +
\frac{1}{s + 1} \sum_{r = 0}^s \delta^a_{a_0} \delta^{b_r}_b
S^{b_0 \cdots b_{r - 1} b_{r + 1} \cdots b_s}_{a_1 \cdots a_s} \Bigr),
\\ 
S^{b_0 \cdots b_s}_{a_0 \cdots a_s} &= \frac{1}{s + 1}
\sum_{r = 0}^s \delta^{b_r}_{a_0}
S^{b_0 \cdots b_{r - 1} b_{r + 1} \cdots b_s}_{a_1 \cdots a_s}. 
\end{align*}

Furthermore, ${\bar\gamma}^B_A = \alpha(\bar{\phi}_A) \delta^B_A$ are 
arbitrary diagonal matrices whose entries $\alpha(\bar{\phi}_A)$, 
in general, may be any (real) numbers. By definition, cf.~Eq.~(\ref{II8}),
$\alpha(\bar{\phi}_A)$ is the (up to now arbitrary) 
Weyl weight of the antifields $\bar\phi$. (This arbitrariness may be traced
back to fact that these representations of $sl(1,2)$ are not completely 
reducible, cf.~\cite{15}). Taking advantage of that freedom we may fix 
$\alpha(\bar{\phi}_A)$ by relating it to the Weyl weight 
$\alpha(\phi^A)$ of the fields $\phi^A$ -- which is uniquely determined 
by means of the quantum master equations at the lowest order of $\hbar$ 
(see Sect. 4 and 6 below) -- according to
\begin{equation}
\label{IIQ}
{\bar\gamma}^B_A + \gamma^B_A + 2 \delta^B_A = 0,
\qquad
{\rm i.e.,}
\qquad
\alpha(\bar{\phi}_A) + \alpha(\phi^A) + 2 =0,
\end{equation}
where $\gamma^B_A = \alpha(\phi^A) \delta^B_A$ is the 
analogous (diagonal) matrix in the $sl(1,2)$--representations of the fields
\footnote{
The requirement (\ref{IIQ}) 
ensures that (proper) solutions $S_m$ of the $m$--extended {\it classical} 
master equations can be constructed having vanishing Weyl weight, 
$\alpha(S_m) = 0$. Later on, we identify the Weyl weight of the (anti)fields
with the new ghost number introduced in Ref. \cite{10}.}.
These matrices $\gamma^B_A $ are given by
\begin{equation}
\label{II10}
\gamma^B_A \equiv \begin{cases} 
\delta^{\beta_s}_{\alpha_s} (s + 2) 
\delta^{b_1}_{a_1} \cdots \delta^{b_s}_{a_s}  
& \text{for $A = \alpha_s|a_1 \cdots a_s, B = \beta_s|b_1 \cdots b_s$},
\\
\delta^{\beta_s}_{\alpha_s} (s + 1) 
\delta^{b_0}_{a_0} \cdots \delta^{b_s}_{a_s}  
& \text{for $A = \alpha_s|a_0 \cdots a_s, B = \beta_s|b_0 \cdots b_s$},
\\
0 & \text{otherwise}.
\end{cases}
\end{equation}
From their entries one may read off 
the Weyl weight $\alpha(\phi^A)$ of the fields $\phi^A$, namely 
\begin{equation}
\label{II11}
\alpha(\phi^A) = ( 0, s + 2, s + 1 ),
\end{equation}
and, throught Eq.~(\ref{IIQ}), the Weyl weights of the 
antifields $\bar{\phi}_A, \phi_{A a}^*$ and $\eta_A$, 
\begin{equation}
\label{II12} 
\alpha(\bar{\phi}_A) = - \alpha(\phi^A) - 2,
\qquad
\alpha(\phi_{A a}^*) = - \alpha(\phi^A) - 1,
\qquad
\alpha(\eta_A) = - \alpha(\phi^A).
\end{equation}  

In order to prove that the transformations (\ref{II7}) and (\ref{II8}) obey 
the $sl(1,2)$--superalgebra one needs the basic properties (\ref{II4}) of the
matrices $\sigma_\alpha$ and the following two equalities:
\begin{align*}
\epsilon^{ac} \delta^b_d + \epsilon^{bc} \delta^a_d &= 
- (\sigma^\alpha)^{ab} (\sigma_\alpha)^c_{~d},
\\
(\sigma^\alpha)^{ab} \bigr(
(\sigma_\alpha)^c_{~e} \delta^d_f + \delta^c_e (\sigma_\alpha)^d_{~f} \bigr)
&=
(\sigma^\alpha)^{ab} \bigr(
(\sigma_\alpha)^d_{~e} \delta^c_f + \delta^d_e (\sigma_\alpha)^c_{~f} \bigr),
\end{align*}
which can be proven by means of the following relations:
\begin{equation*}
\epsilon^{ab} \delta^c_d + 
\epsilon^{bc} \delta^a_d + 
\epsilon^{ca} \delta^b_d = 0,
\qquad
\epsilon^{ab} ( \delta^c_e \delta^d_f - \delta^d_e \delta^c_f ) =
\epsilon^{cd} ( \delta^a_e \delta^b_f - \delta^b_e \delta^a_f ).
\end{equation*}


\section{Superspace representations of the algebra $sl(1,2)$}
\setcounter{equation}{0}


This Section is devoted to a geometric interpretation of
the superalgebra $sl(1,2)$ as given by Eqs. (\ref{II6}). This opens
the possibility to formulate the quantization of general gauge theories
in terms of super(anti)fields over a 2--dimensional superspace.

\medskip
\noindent (A) {\it Representations of sl(1,2) in superspace}
\\
In Ref. \cite{11} it was pointed out that the generators of the (real)
algebra $osp(1,1|2) \sim sl(1,2)$ acquire a clear geometric meaning if 
they are interpreted as generators of transformations in superspace.
This is obtained by redefining the generators of $sl(1,2)$ as follows: 
\begin{equation}
\label{III13}
V_+^a \equiv - i P^a,
\qquad
V_-^a \equiv - i K^a,
\qquad
V^{ab} \equiv - i M^{ab},
\qquad
V \equiv i D.
\end{equation}
Then, the (anti)commutation relations resulting from (\ref{II6}) can be
interpreted as algebra of the conformal group in two {\em anticommuting}
dimensions with metric tensor $\epsilon^{ab}$: 
\begin{alignat}{3}
\hspace{-.2cm}
[ D, M^{ab} ] &= 0,
&\quad
[ D, P^a ] &= - i P^a,
&\quad
[ D, K^a ] &= i K^a,
\nonumber
\\
\label{III14}
\hspace{-.2cm}
[ M^{ab}, M^{cd} ] &= - i \epsilon^{\{c \{a } M^{ b\} d\} }, 
&\quad 
[ M^{ab}, P^c ] &= - i \epsilon^{c \{a } P^{ b\} }, 
&\quad 
[ M^{ab}, K^c ] &= - i \epsilon^{c \{a } K^{ b\} }, 
\\ 
\hspace{-.2cm}
\{ P^a, P^b \} &= 0,
&\quad
\{ K^a, K^b \} &= 0,
&\quad
\{ P^a, K^b \} &= i ( \epsilon^{ab} D - M^{ab} ), 
\nonumber
\end{alignat}
with $P^a$, $K^a$, $M^{ab}$ and $D$ being the generators of translations, 
special conformal transformations, (symplectic) rotations and 
dilatations, respectively. The superspace which we encounter here is obtained
by extending the usual spacetime to include two extra anticommuting 
coordinates $\theta^a$. Raising and lowering of $Sp(2)$--indices are 
defined by the rules $\theta^a = \epsilon^{ab} \theta_b$ and 
$\theta_a = \epsilon_{ab} \theta^b$; the square of $\theta^a$ and the 
derivative with respect to it are defined by
$\theta^2 \equiv \hbox{$\frac{1}{2}$} \epsilon_{ab} \theta^b \theta^a$ and
$\partial^2/ \partial \theta^2 \equiv \hbox{$\frac{1}{2}$} \epsilon^{ab}
\partial^2/ \partial \theta^b \partial \theta^a$.

The representation of the algebra (\ref{III14}) in that superspace is given by
\begin{align}
\label{III15}
P^a &= i \frac{\partial}{\partial \theta_a},
\\
\label{III16}
K^a &= 2 i \theta^2 \frac{\partial}{\partial \theta_a} - 
\theta_b ( \Sigma^{ab} - i \epsilon^{ab} \Delta ),
\\
\label{III17}
M^{ab} &= - i \Bigr( 
\theta^a \frac{\partial}{\partial \theta_b} +
\theta^b \frac{\partial}{\partial \theta_a} \Bigr) + \Sigma^{ab},
\\
\label{III18}
D &= i \theta_a \frac{\partial}{\partial \theta_a} - i \Delta,
\end{align}
where $\Sigma^{ab}$ and $\Delta$ constitute the basis of some 
finite--dimensional representation of the algebra of the ``little group'', 
i.e., the stabilizer subgroup of that conformal group, 
\begin{equation*}
[ \Sigma^{ab}, \Sigma^{cd} ] = - \epsilon^{ \{c \{a } \Sigma^{ b\} d\} },
\qquad
[ \Delta, \Sigma^{ab} ] = 0.
\end{equation*}
Obviously, the corresponding representation of the algebra (\ref{II6}) is 
obtained by a change of the $SL(2)$--generators analogous to (\ref{II5}),
$\Sigma^{ab} = i (\sigma^\alpha)^{ab} \Sigma_\alpha$, with $\Sigma_\alpha$ 
being related to the matrix representation of the $V_\alpha$'s and 
satisfying
\begin{equation}
\label{III19}
[ \Sigma_\alpha, \Sigma_\beta ] = \epsilon_{\alpha\beta}^{~~~\!\gamma}
\Sigma_\gamma,
\qquad
[ \Delta, \Sigma_\alpha ] = 0.
\end{equation}
The corresponding representation of the generators (\ref{III13}) in the 
superspace are
\begin{align}
\label{III15a}
V_+^a &= \frac{\partial}{\partial \theta_a},
\\
\label{III16a}
V_-^a &= 2 \theta^2 \frac{\partial}{\partial \theta_a} - 
\theta_b \bigr( (\sigma^\alpha)^{ab} \Sigma_\alpha - 
\epsilon^{ab} \Delta \bigr),
\\
\label{III17a}
V_\alpha &= \theta^a (\sigma_\alpha)^a_{~b}
\frac{\partial}{\partial \theta_b} + \Sigma_\alpha,
\\
\label{III18a}
V &= - \theta_a \frac{\partial}{\partial \theta_a} + \Delta.
\end{align}
\smallskip
\noindent (B) {\it Representation of $sl(1,2)$ on super(anti)fields}
\\
Now, having revealed the geometrical content of the generators of $sl(1,2)$ 
we are able to formulate the transformations (\ref{II7}) and (\ref{II8})
in superspace. Let $\Phi^A(\theta)$, 
$\epsilon(\Phi^A(\theta)) \equiv \epsilon_A$, be a set of superfields with 
the restriction $\Phi^A(\theta)|_{\theta = 0} = \phi^A$. It admits the 
following general expansion in terms of component fields,
\begin{equation}
\label{III24}
\Phi^A(\theta) = \phi^A + \pi^{A a} \theta_a - \lambda^A \theta^2,
\qquad
\frac{\delta}{\delta \Phi^A(\theta)} = \frac{\delta}{\delta \phi^A} \theta^2 -
\theta^a \frac{\delta}{\delta \pi^{A a}} -
\frac{\delta}{\delta \lambda^A}
\end{equation} 
(remember that, according to the general convention, derivatives 
with respect to the fields are defined as acting from the {\em right}).
With each superfield $\Phi^A(\theta)$ a superantifield 
$\bar{\Phi}_A(\theta)$ is associated having the {\em same} Grassmann parity,
$\epsilon(\bar{\Phi}_A(\theta)) = \epsilon_A$,
\begin{equation}
\label{III25}
\bar{\Phi}_A(\theta) = \bar{\phi}_A - \theta^a \phi^*_{A a} -
\theta^2 \eta_A,
\qquad
\frac{\delta}{\delta \bar{\Phi}_A(\theta)} =
\theta^2 \frac{\delta}{\delta \bar{\phi}_A} +
\frac{\delta}{\delta \phi^*_{A a}} \theta_a -
\frac{\delta}{\delta \eta_A}.
\end{equation}
According to (\ref{III24}) and (\ref{III25}) for the expressions of the
derivatives it holds
\begin{equation*}
\frac{\delta \Phi^A(\theta)}{\delta \Phi^B(\bar{\theta})} =
\frac{\delta \bar{\Phi}_B(\theta)}{\delta \bar{\Phi}_A(\bar{\theta})} =
\delta^A_B \delta^2(\theta - \bar{\theta}), 
\qquad
\hbox{with}
\qquad
\delta^2(\theta - \bar{\theta}) \equiv (\theta - \bar{\theta})^2.
\end{equation*}
Then, by the help of $\bar{\Phi}_A(\theta)$ the $sl(1,2)$-transformations 
(\ref{II7}) and (\ref{II8}) may be written in the following compact form:
\begin{align}
\label{III26}
V_+^a \bar{\Phi}_A(\theta) &= 
\frac{\partial \bar{\Phi}_A(\theta)}{\partial \theta_a},
\\
\label{III27}
V_-^a \bar{\Phi}_A(\theta) &= 
2 \theta^2 \frac{\partial \bar{\Phi}_A(\theta)}{\partial \theta_a} - 
\theta_b \bigr(
(\sigma^\alpha)^{ab} \Sigma_\alpha - \epsilon^{ab} \Delta \bigr)
\bar{\Phi}_A(\theta),
\\
\label{III28}
V_\alpha \bar{\Phi}_A(\theta) &= - \Bigr\{
\theta_a (\sigma_\alpha)^a_{~b}
\frac{\partial \bar{\Phi}_A(\theta)}{\partial \theta_b} + 
\Sigma_\alpha \bar{\Phi}_A(\theta) \Bigr\},
\\
\label{III29}
V \bar{\Phi}_A(\theta) &= - \Bigr\{
- \theta_a \frac{\partial \bar{\Phi}_A(\theta)}{\partial \theta_a} + 
\Delta \bar{\Phi}_A(\theta) \Bigr\}
\end{align}
with
\begin{equation}
\label{III30}
\Sigma_\alpha \bar{\Phi}_A(\theta) = - \bar{\Phi}_B(\theta)
(\sigma_\alpha)^B_{~~\!A},
\qquad
\Delta \bar{\Phi}_A(\theta) = - \bar{\Phi}_B(\theta) 
{\bar\gamma}^B_{~~\!A}.
\end{equation}
Some care has to be taken in order to get the correct signs in these 
equations. First, in order to attain that the transformations laws
(\ref{III26})--(\ref{III29}) are compatible with the superalgebra (\ref{II6})
it is necessary to take into account an extra minus sign on the right--hand 
side of (\ref{III28}) and (\ref{III29}) 
(cf. Eqs. (\ref{III17a}), (\ref{III18a})).
Since the matrices $\Sigma_\alpha$ generate an irreducible representation of 
the symplectic group, by virtue of (\ref{III19}), $- \Delta$ must be a number 
which, by definition, agrees with the Weyl weight of the superantifields 
(observe $\alpha(\theta)=1$ in accordance with Eqs. (\ref{II12})). 
Secondly, let us emphasize that the minus sign on the 
right--hand side of the first relation (\ref{III30}) is crucial: 
A further transformation in (\ref{III28}) does not act on the 
numerical matrices $\Sigma_\alpha$ but directly on $\bar{\Phi}_A(\theta)$; 
this reverses the factors on the right--hand side against those on the left 
one and the minus sign is therefore necessary to retain the multiplication 
law of the conformal group. 

Collecting the results obtained up to now the representation of the generators 
of $sl(1,2)$ by differential operators on the superspace reads
\begin{align}
\label{III31}
V_+^a &= \int d^2 \theta \,
\frac{\partial \bar{\Phi}_A(\theta)}{\partial \theta_a}
\frac{\delta}{\delta \bar{\Phi}_A(\theta)},
\\
\label{III32}
V_-^a &= \int d^2 \theta \, \Bigr\{
2 \theta^2 \frac{\partial \bar{\Phi}_A(\theta)}{\partial \theta_a} + 
\theta_b \bar{\Phi}_B(\theta) \bigr(
(\sigma^\alpha)^{ab} (\sigma_\alpha)^B_{~~\!A} -
\epsilon^{ab} {\bar\gamma}^B_A \bigr) \Bigr\}
\frac{\delta}{\delta \bar{\Phi}_A(\theta)},
\\
\label{III33}
V_\alpha &= \int d^2 \theta \, \Bigr\{
- \theta_a (\sigma_\alpha)^a_{~b}
\frac{\partial \bar{\Phi}_A(\theta)}{\partial \theta_b} + 
\bar{\Phi}_B(\theta) (\sigma_\alpha)^B_{~~\!A} \Bigr\}
\frac{\delta}{\delta \bar{\Phi}_A(\theta)},
\\
\label{III34}
V &= \int d^2 \theta \, \Bigr\{
\theta_a \frac{\partial \bar{\Phi}_A(\theta)}{\partial \theta_a} + 
\bar{\Phi}_B(\theta) {\bar\gamma}^B_A \Bigr\}
\frac{\delta}{\delta \bar{\Phi}_A(\theta)},
\end{align}
where the integration over $\theta^a$ is given by
\begin{equation*}
\int d^2 \theta = 0,
\qquad
\int d^2 \theta \, \theta^a = 0,
\qquad
\int d^2 \theta \, \theta^a \theta^b = \epsilon^{ab}.
\end{equation*}
Making use of the expansions (\ref{III25}) for $\bar{\Phi}_A(\theta)$
and $\delta/ \delta \bar{\Phi}_A(\theta)$ and performing in 
Eqs. (\ref{III31})--(\ref{III34}) the $\theta$--integration 
it is easily verified that the resulting expressions for $V_\pm^a$, 
$V_\alpha$ and $V$ generate exactly the transformations 
(\ref{II7}) and (\ref{II8}) of the component fields of $\bar{\Phi}_A(\theta)$.

\smallskip
Furthermore, let us give also a superspace representation of $sl(1,2)$ in 
terms of $\Phi^A(\theta)$. The corresponding generators $U_\pm^a$, $U_\alpha$ 
and $U$ being defined as {\it right} derivatives -- in contrast to 
$V_\pm^a$, $V_\alpha$ and $V$, which are defined as {\it left} ones -- 
obey the following (anti)commutation relations (cf. Eqs. (\ref{II3}))
\begin{alignat}{3}
[ U, U_\alpha ] &= 0,
&\qquad
[ U, U_+^a ] &= - U_+^a,
&\qquad
[ U, U_-^a ] &= U_-^a,
\nonumber
\\
\label{III35}
[ U_\alpha, U_\beta ] &= - \epsilon_{\alpha\beta}^{~~~\!\gamma} U_\gamma, 
&\qquad 
[ U_\alpha, U_+^a ] &= - U_+^b (\sigma_\alpha)_b^{~a}, 
&\qquad 
[ U_\alpha, U_-^a ] &= - U_-^b (\sigma_\alpha)_b^{~a}, 
\\ 
\{ U_+^a, U_+^b \} &= 0,
&\qquad
\{ U_-^a, U_-^b \} &= 0,
&\qquad
\{ U_+^a, U_-^b \} &= (\sigma^\alpha)^{ab} U_\alpha + \epsilon^{ab} U.
\nonumber
\end{alignat}
If we replace in Eqs. (\ref{III31})--(\ref{III34}) the superantifield
$\bar{\Phi}_A(\theta)$ by $\Phi^A(\theta)$, the left derivatives
$\delta_L/ \delta \bar{\Phi}_A(\theta)$ by the right derivatives 
$\delta_R/ \delta \Phi^A(\theta)$, and reverse the order of all the factors, 
then for the representations we are looking for we obtain
\begin{align}
\label{III36}
U_+^a &= \int d^2 \theta \, \frac{\delta}{\delta \Phi^A(\theta)}
\frac{\partial_R \Phi^A(\theta)}{\partial \theta_a},
\\
\label{III37}
U_-^a &= \int d^2 \theta \, \frac{\delta}{\delta \Phi^A(\theta)} \Bigr\{
2 \theta^2 \frac{\partial_R \Phi^A(\theta)}{\partial \theta_a} + 
\bigr( (\sigma^\alpha)^{ab} (\sigma_\alpha)^A_{~~\!B} + 
\epsilon^{ab} \gamma^A_B \bigr)  
\Phi^B(\theta) \theta_b \Bigr\},
\\
\label{III38}
U_\alpha &= \int d^2 \theta \, \frac{\delta}{\delta \Phi^A(\theta)} \Bigr\{
- \frac{\partial_R \Phi^A(\theta)}{\partial \theta_b}
(\sigma_\alpha)_b^{~a} \theta_a + 
(\sigma_\alpha)^A_{~~\!B} \Phi^B(\theta) \Bigr\},
\\
\label{III39}
U &= \int d^2 \theta \,
\frac{\delta}{\delta \Phi^A(\theta)} \Bigr\{
\frac{\partial_R \Phi^A(\theta)}{\partial \theta_a} \theta_a + 
\gamma^A_B \Phi^B(\theta) \Bigr\}.
\end{align}
In addition, we have replaced ${\bar\gamma}^B_A$ by the (diagonal)
matrix $\gamma^B_A = \alpha(\phi^A) \delta^B_A$, whose entries 
$\alpha(\phi^A)$ are given by Eq. (\ref{II11}).

Making use of the expansions (\ref{III25}) for $\Phi^A(\theta)$ and
$\delta/\delta \Phi^A(\theta)$ and integrating in 
Eqs. (\ref{III36})--(\ref{III39}) over $\theta^a$ for the components of 
$\Phi^A(\theta)$ one obtains the (linear) transformations 
\begin{alignat}{2}
\phi^A U_+^a &= \pi^{A a},
&\quad
\phi^A U_-^a &= 0,
\nonumber
\\
\label{U1}
\pi^{A b} U_+^a &= - \epsilon^{ab} \lambda^A,
&\quad
\pi^{A b} U_-^a &= \bigr(
(\sigma^\alpha)^{ab} (\sigma_\alpha)^A_{~~\!B} + 
\epsilon^{ab} \gamma^A_B \bigr) \phi^B,
\\
\lambda^A U_+^a &= 0,
&\quad
\lambda^A U_-^a &= \bigr(
(\sigma^\alpha)^a_{~b} (\sigma_\alpha)^A_{~~\!B} + 
\delta^a_b ( \gamma^A_B + 2 \delta^A_B ) \bigr) \pi^{B b},
\nonumber
\\
\nonumber
\\
\phi^A U_\alpha &= (\sigma_\alpha)^A_{~~\!B} \phi^B,
&\quad
\phi^A U &= \gamma^A_B \phi^B,
\nonumber
\\
\label{U2}
\pi^{A a} U_\alpha &= (\sigma_\alpha)^A_{~~\!B} \pi^{B a} +
(\sigma_\alpha)^a_{~b} \pi^{A b},
&\quad
\pi^{A a} U &= ( \gamma^A_B + \delta^A_B ) \pi^{B a},
\\
\lambda^A U_\alpha &= (\sigma_\alpha)^A_{~~\!B} \lambda^B,
&\quad
\lambda^A U &= ( \gamma^A_B + 2 \delta^A_B ) \lambda^B,
\nonumber
\end{alignat}
which define the explicit realization of $sl(1,2)$ on the superfield
analogous to Eqs. (\ref{II7}) and (\ref{II8}). By a simple straightforward 
calculation it is verified that the transformations (\ref{U1}) and (\ref{U2})
indeed satisfy the $sl(1,2)$--superalgebra (\ref{III35}). 


\section{Quantum master equations}
\setcounter{equation}{0}


The superspace representation of $sl(1,2)$ obtained in the previous section
enables one to attack the problem of superfield quantization of general 
gauge theories.  
A superfield version for the $Sp(2)$--covariant Lagrangian quantization 
was proposed in Ref. \cite{4}. In that approach the quantum action
$W(\Phi^A(\theta), \bar{\Phi}_A(\theta))$ is required to be invariant under 
the (anti)BRST transformations which, in superspace, are realized as 
translations along the coordinates $\theta^a$. 

In order to proceed further in the development of that formalism one may 
attempt 
to include also special conformal transformations, symplectic rotations and 
dilatations by imposing additional symmetry requirements. Such an 
extension is possible, but only for one of the two 
$osp(1,2)$--subalgebras of $osp(2,2) \sim sl(1,2)$. Indeed, for a 
superfield description of the $osp(1,2)$--covariant quantization procedure 
introduced in Ref. \cite{6} one needs both translations as well as 
special conformal transformations and symplectic rotations. In that
approach the translations are combined with the special conformal
transformations by means of a mass parameter $m$ leading to $m$--dependent
(anti)BRST transformations. The invariance under symplectic transformations
ensures the ghost number conservation of the corresponding quantum 
action $W_m(\Phi^A(\theta), \bar{\Phi}_A(\theta))$. In addition, the
dilatations may be used to ensure the new ghost number conservation of 
$W_m(\Phi^A(\theta), \bar{\Phi}_A(\theta))$ at the lowest order of $\hbar$. 

\smallskip
\noindent (A) {\it Sp(2)--covariant superfield quantization }
\\
To begin with, we shortly review the $Sp(2)$--covariant superfield
quantization \cite{4}. Let us introduce the antisymplectic differential 
operators
\begin{equation}
\label{IV40}
\bar{\Delta}^a = \Delta^a + (i/\hbar) V^a,
\qquad
V^a \equiv V_+^a,
\end{equation}
with the translations $V_+^a$ given by Eq.~(\ref{III31}) and
the nilpotent (second--order) differential operators 
$\Delta^a$ given by
\begin{equation}
\label{Delta}
\Delta^a = \int d^2 \theta 
\frac{\partial^2 \delta_L}{\partial \theta^2 \delta \Phi^A(\theta)} \,
\theta^a \frac{\delta}{\delta \bar{\Phi}_A(\theta)} = 
(-1)^{\epsilon_A} \frac{\delta_L}{ \delta \phi^A} \, 
\frac{\delta}{ \delta \phi_{A a}^*}.
\end{equation}
Let us remark, that this definition of $\Delta^a$ by projecting out from 
$\delta_L/ \delta \Phi^A(\theta)$ only the first component agrees with the 
initial definition in Ref. \cite{10} but differs from that in Ref. \cite{4}. 
In our opinion the definition (\ref{Delta}) seems to be much better adapted 
to the present aim than that of Ref. \cite{4} since a change of the definition 
of $\Delta^a$, like in the triplectic quantization \cite{16}, requires also 
a change of the definition of $V^a$ -- but then the geometric meaning of 
$V^a$ would be lost. The operators $\bar{\Delta}^a$, $\Delta^a$ and $V^a$
possess the important properties of nilpotency 
and (relative) anticommutativity,
\begin{equation*}
\{ \bar{\Delta}^a, \bar{\Delta}^b \} = 0
\quad
\Longleftrightarrow
\quad
\{ \Delta^a, \Delta^b \} = 0,
\quad
\{ V^a, V^b \} = 0,
\quad
\{ \Delta^a, V^b \} = 0.
\end{equation*}

The basic object of the superfield quantization is the quantum action
$W(\Phi^A(\theta), \bar{\Phi}_A(\theta))$, which is required to be a solution
of the quantum master equation
\begin{equation}
\label{IV41}
\bar{\Delta}^a\, {\rm exp}\{ (i/ \hbar) W \} = 0
\qquad
\Longleftrightarrow 
\qquad
\hbox{$\frac{1}{2}$} ( W, W )^a + V^a W = i \hbar \Delta^a W,
\end{equation}
where the superantibrackets $( F,G )^a$ are defined by 
\begin{equation}
\label{IV42}
( F,G )^a = (-1)^{\epsilon_A} \int d^2 \theta \, \Bigr\{ 
\frac{\partial^2 \delta F}
{\partial \theta^2 \delta \Phi^A(\theta)} \theta^a 
\frac{\delta G}{\delta \bar{\Phi}_A(\theta)} - 
(-1)^{(\epsilon(F) + 1) (\epsilon(G) + 1)} (F \leftrightarrow G) \Bigr\}. 
\end{equation}
The solution of (\ref{IV41}) is sought of as a power series in Planck's
constant $\hbar$, 
\begin{equation*}
W = S + \sum_{n = 1}^\infty \hbar^n W_{n}.
\end{equation*}
Furthermore, two requirements -- the nondegeneracy of $S$ 
and the correctness of the classical limit -- have to be imposed. The first 
one is translated into the requirement that $S$ should be a 
{\em proper} solution of the classical master equation, i.e., 
the Hessian  of second derivatives of $S$ 
should be of maximal rank at the stationary points, and the second one 
means that $S$ should satisfy the usual boundary condition, namely 
that $S$ coincides with the classical action $S_{\rm cl}(A)$ if all the
antifields are put equal to zero.

To remove the gauge degeneracy of the action $S$, one introduces the operator
\begin{equation*}
\hat{U}(F) = {\rm exp}\{ (\hbar/ i) \hat{T}(F) \}
\quad
{\rm with}
\quad
\hat{T}(F) = \hbox{$\frac{1}{2}$} \epsilon_{ab} 
\{ \bar{\Delta}^b, [ \bar{\Delta}^a, F ] \},
\end{equation*}
$F = F(\Phi^A(\theta))$ being an arbitrary bosonic gauge fixing 
functional. Then, the gauge fixed quantum action 
$W_{\rm ext}(\Phi^A(\theta), \bar{\Phi}_A(\theta))$, defined by  
\begin{equation}
\label{IV43}
{\rm exp}\{ (i/ \hbar) W_{\rm ext} \} = 
\hat{U}(F) \,{\rm exp}\{ (i/ \hbar) W \}, 
\end{equation}
is also a solution of the quantum master equations (\ref{IV41}).

\smallskip
\noindent (B) {\it osp(1,2)--covariant superfield quantization } 
\\
Let us now give the superfield description of the $osp(1,2)$--covariant 
quantization \cite{6}. In that approach the antisymplectic differential
operators (\ref{IV40}) are replaced by
\begin{equation}
\label{IV44}
\bar{\Delta}_m^a = \Delta^a + (i/\hbar) V_m^a,
\qquad
V_m^a \equiv V_+^a + \hbox{$\frac{1}{2}$} m^2 V_-^a,
\end{equation}
with the special conformal operators $V_-^a$ given by Eq. (\ref{III32}).
Here, the mass parameter $m$ having Weyl weight $\alpha(m) = 1$ is introduced 
because $V_+^a$ and $V_-^a$ have different mass dimensions (and opposite 
Weyl weight $\alpha(V_\pm^a) = \pm 1$). In addition, one introduces the 
differential operators
\begin{equation}
\label{IV45}
\bar{\Delta}_\alpha = \Delta_\alpha + (i/\hbar) V_\alpha,
\end{equation}
with the symplectic rotations $V_\alpha$ given by Eq. (\ref{III33}) and 
the (second--order) differential operators $\Delta_\alpha$ being defined by 
\begin{equation*}
\Delta_\alpha = (-1)^{\epsilon_A + 1} \int d^2 \theta \,  
\theta^2 (\sigma_\alpha)_B^{~~\!A} \frac{\partial^2 \delta_L}
{\partial \theta^2 \delta \Phi^A(\theta)} \,
\frac{\delta}{\delta \bar{\Phi}_B(\theta)} =
(-1)^{\epsilon_A} (\sigma_\alpha)_B^{~~\!A} 
\frac{\delta_L}{ \delta \phi^A} \frac{\delta}{\delta \eta_B}.
\end{equation*}
As long as $m \neq 0$ the operators $\bar{\Delta}_m^a$ are neither nilpotent
nor do they anticommute among themselves; instead, together with the
operators $\bar{\Delta}_\alpha$ they generate a superalgebra isomorphic
to $osp(1,2)$:
\begin{alignat}{2}
[ V_\alpha, V_\beta ] &= \epsilon_{\alpha\beta}^{~~~\!\gamma} V_\gamma, 
&\qquad\qquad
[ \bar{\Delta}_\alpha, \bar{\Delta}_\beta ] &= (i/\hbar) 
\epsilon_{\alpha\beta}^{~~~\!\gamma} \bar{\Delta}_\gamma, 
\nonumber
\\
\label{IV46}
[ V_\alpha, V_m^a ] &= V_m^b (\sigma_\alpha)_b^{~a}, 
&\qquad\qquad
[ \bar{\Delta}_\alpha, \bar{\Delta}_m^a ] &= (i/\hbar)
\bar{\Delta}_m^b (\sigma_\alpha)_b^{~a}, 
\\
\{ V_m^a, V_m^b \} &= - m^2 (\sigma^\alpha)^{ab} V_\alpha, 
&\qquad\qquad
\{ \bar{\Delta}_m^a, \bar{\Delta}_m^b \} &= - (i/\hbar)
m^2 (\sigma^\alpha)^{ab} \bar{\Delta}_\alpha. 
\nonumber
\end{alignat}

The $m$--{\it dependent} quantum action 
$W_m(\Phi^A(\theta), \bar{\Phi}_A(\theta))$ is required to obey the 
$m$--{\it extended} generalized quantum master equations
\begin{align}
\label{IV47}
\bar{\Delta}_m^a\, {\rm exp}\{ (i/ \hbar) W_m \} = 0
\qquad
&\Longleftrightarrow 
\qquad
\hbox{$\frac{1}{2}$} ( W_m, W_m )^a + V_m^a W_m = i \hbar \Delta^a W_m
\\
\intertext{which ensure (anti)BRST invariance, and the generating equations 
of $Sp(2)$--invariance:}
\label{IV48}
\bar{\Delta}_\alpha\, {\rm exp}\{ (i/ \hbar) W_m \} = 0
\qquad
&\Longleftrightarrow 
\qquad
\hbox{$\frac{1}{2}$} \{ W_m, W_m \}_\alpha + V_\alpha W_m =
i \hbar \Delta_\alpha W_m,
\end{align}
where the curly superbrackets $\{ F,G \}_\alpha$ are defined by 
\begin{equation}
\label{IV49}
\{ F,G \}_\alpha = - \int d^2 \theta \, \Bigr\{ 
\theta^2 \frac{\partial^2 \delta F}
{\partial \theta^2 \delta \Phi^A(\theta)} \,
\frac{\delta G}{\delta \bar{\Phi}_B(\theta)} (\sigma_\alpha)_B^{~~\!A} +
(-1)^{\epsilon(F) \epsilon(G)} (F \leftrightarrow G) \Bigr\}.
\end{equation}

The gauge fixed quantum action 
$W_{m, {\rm ext}}(\Phi^A(\theta), \bar{\Phi}_A(\theta))$ is introduced
according to 
\begin{equation}
\label{IV50}
{\rm exp}\{ (i/ \hbar) W_{m, {\rm ext}} \} = 
\hat{U}_m(F) \,{\rm exp}\{ (i/ \hbar) W_m \}, 
\end{equation}
where the operator $\hat{U}_m(F)$ has to be choosen as \cite{6}
\begin{equation*}
\hat{U}_m(F) = {\rm exp}\{(\hbar/ i) \hat{T}_m(F)\}
\quad
{\rm with}
\quad
\hat{T}_m(F) = \hbox{$\frac{1}{2}$} \epsilon_{ab} 
\{ \bar{\Delta}_m^b, [ \bar{\Delta}_m^a, F ] \} + (i/ \hbar)^2 m^2 F,
\end{equation*}
$F = F(\Phi^A(\theta))$ being the gauge fixing functional. With these
definitions one establishes the following two relations:
\begin{equation*}
[ \bar{\Delta}_m^a, \hat{T}_m(F) ] = \hbox{$\frac{1}{2}$} (i/\hbar)
(\sigma^\alpha)^a_{~b} [ \bar{\Delta}_m^b, [ \bar{\Delta}_\alpha , F ] ]
\end{equation*}
\vspace*{-1cm}
\begin{equation*}
[ \bar{\Delta}_\alpha, \hat{T}_m(F) ] = \hbox{$\frac{1}{2}$} 
\epsilon_{ab} \left\{ \bar{\Delta}_m^b, 
[ \bar{\Delta}_m^a, [ \bar{\Delta}_\alpha, F ] ] \right\} +
(i/\hbar)^2 m^2 [ \bar{\Delta}_\alpha, F ].
\end{equation*}
Restricting $F(\Phi^A(\theta))$ to be a $Sp(2)$--scalar by imposing
the condition $[ \bar{\Delta}_\alpha, F ] W_m = 0$ it can be verified 
(see Ref.~\cite{6}) that the commutators 
$[ \bar{\Delta}_m^a, \hat{U}_m(F) ]$ and 
$[ \bar{\Delta}_\alpha, \hat{U}_m(F) ]$, if applied on
${\rm exp}\{ (i/\hbar) W_m \}$, vanish on the subspace of {\em admissible}
 actions $W_m$. These action are determined by the condition 
\begin{equation}
\label{IV51}
\int d^2 \theta \, \theta^2 \Bigr\{
\frac{\delta W_m}{\delta \bar{\Phi}_A(\theta)} + \Phi^A(\theta) \Bigr\} = 0
\qquad
\Longleftrightarrow
\qquad
\frac{\delta W_m}{\delta \eta_A} = \phi^A,  
\end{equation}
i.e., depending only {\it linearly} on $\eta_A$. This condition ensures
that the gauge fixed quantum action $W_{m, {\rm ext}}$ also satisfies the 
quantum master equations (\ref{IV47}) and (\ref{IV48}). Then, by virtue of 
(\ref{IV51}), the restriction $[ \bar{\Delta}_\alpha, F ] W_m = 0$ becomes
\begin{equation*}
[ \bar{\Delta}_\alpha, F ] W_m = 0
\qquad
\Longrightarrow
\qquad
\int d^2 \theta \, \theta^2 \frac{\partial^2 \delta F}
{\partial \theta^2 \delta \Phi^A(\theta)} \, 
\Phi^B(\theta) (\sigma_\alpha)_B^{~~\!A} + V_\alpha F = 0,
\end{equation*}
which expresses the $Sp(2)$--invariance of $F$. Furthermore, the 
quantum master equations (\ref{IV48}) simplify into
\begin{equation}
\label{IV52}
\bar{\Delta}_\alpha {\rm exp}\{ (i/\hbar) W_m \} = 0
\quad
\Longrightarrow
\quad
\int d^2 \theta \, \theta^2 \frac{\partial^2 \delta W_m}
{\partial \theta^2 \delta \Phi^A(\theta)} \, 
\Phi^B(\theta) (\sigma_\alpha)_B^{~~\!A} + V_\alpha W_m = 0,
\end{equation}
since the $\sigma_\alpha$--matrices are traceless. 

The equations (\ref{IV52}) for $\alpha = 0$ express the
ghost number conservation of the action $W_m$, ${\rm gh}(W_m) = 0$. 
Thereby the ghost numbers of the fields and antifields are given by
\begin{align*}
{\rm gh}(\phi^A) &= 
- \Bigr( 0,  \sum_{r = 1}^s (-1)^{a_r},  \sum_{r = 0}^s (-1)^{a_r} \Bigr),
\qquad
\hbox{where} 
\qquad 
a_r = 1,2,
\\
{\rm gh}(\bar{\phi}_A) &= - {\rm gh}(\phi^A),
\qquad
{\rm gh}(\phi_{A a}^*) = - {\rm gh}(\phi^A) + (-1)^a,
\qquad
{\rm gh}(\eta_A) = - {\rm gh}(\phi^A).
\end{align*}

\smallskip
\noindent (C) {\it New ghost number conservation }
\\
In Ref. \cite{10} also a so--called new ghost number was ascribed to all 
fields and antifields of the solutions of the {\it classical} master 
equations in the following way:
\begin{align*}
{\rm ngh}(\phi^A) &= ( 0, s + 2, s + 1 ),
\\
{\rm ngh}(\bar{\phi}_A) = - {\rm ngh}(\phi^A) - 2,
\qquad 
{\rm ngh}(\phi_{A a}^*) &= - {\rm ngh}(\phi^A) - 1,
\qquad
{\rm ngh}(\eta_A) = - {\rm ngh}(\phi^A).
\end{align*}
According to these definitions we also have ${\rm ngh}(\theta^a) = -1$.
In comparison with Eqs. (\ref{II11}) and (\ref{II12}) it follows that 
the new ghost number agrees with the Weyl weight of the fields and 
antifields, i.e., 
\begin{alignat*}{3}
{\rm ngh}(\phi^A) &= \alpha(\phi^A),
&\qquad
{\rm ngh}(\pi^{Aa}) &= \alpha(\pi^{Aa}),
&\qquad
{\rm ngh}(\lambda^A) &= \alpha(\lambda^A),
\\
{\rm ngh}(\bar{\phi}_A) &= \alpha(\bar{\phi}_A),
&\qquad 
{\rm ngh}(\phi_{A a}^*) &= \alpha(\phi_{A a}^*),
&\qquad
{\rm ngh}(\eta_A) &= \alpha(\eta_A).
\end{alignat*}

In order to clarify how in our approach both numbers are related 
to each other let us introduce the following differential operator
\begin{equation}
\label{IV54}
\bar{\Delta}_m = \Delta + (i/\hbar) V_m,
\qquad
V_m \equiv V + m \frac{\partial}{\partial m},
\end{equation}
with the dilatations $V$ given by Eq. (\ref{III34}) and the (second--order)
differential operator $\Delta$ defined by
\begin{equation*}
\Delta = (-1)^{\epsilon_A + 1} \int d^2 \theta \,  
\theta^2 \gamma_B^A \frac{\partial^2 \delta_L}
{\partial \theta^2\delta \Phi^A(\theta)} 
\frac{\delta}{\delta \bar{\Phi}_B(\theta)} = 
(-1)^{\epsilon_A}  \gamma_B^A \frac{\delta_L}{ \delta \phi^A} 
\frac{\delta}{\delta \eta_B}.
\end{equation*}
The new operator $\bar{\Delta}_m$ together with the generating operators 
$\bar{\Delta}_m^a$ and $\bar{\Delta}_\alpha$ form an extension of the 
$osp(1,2)$--superalgebra being isomorphic to $osp(1,2) \oplus u(1)$ where,
in addition to the (anti)commutation relations (\ref{IV46}), the following
relations hold true:
\begin{alignat}{2}
[ V_m, V_m ] &= 0,
&\qquad
[ \bar{\Delta}_m, \bar{\Delta}_m ] &= 0,
\nonumber
\\
\label{Delta_m}
[ V_m, V_\alpha ] &= 0,
&\qquad
[ \bar{\Delta}_m, \bar{\Delta}_\alpha ] &= 0,
\\
[ V_m, V_m^a ] &= V_m^a,
&\qquad
[ \bar{\Delta}_m, \bar{\Delta}_m^a ] &= (i/\hbar) \bar{\Delta}_m^a.
\nonumber
\end{alignat}

Let us assume now that solutions $W_m$ of the quantum master equations 
(\ref{IV48}) and (\ref{IV49}) can be constructed which also satisfy the 
following equation: 
\begin{equation}
\label{IV55}
\bar{\Delta}_m\, {\rm exp}\{ (i/ \hbar) W_m \} = 0
\qquad
\Longleftrightarrow 
\qquad
\hbox{$\frac{1}{2}$} \{ W_m, W_m \} + V_m W_m = i \hbar \Delta_m W_m
\end{equation}
with the following abbreviation
\begin{equation}
\label{IV56}
\{ F,G \} = - \int d^2 \theta \, \Bigr\{ 
\theta^2 \frac{\partial^2 \delta F}
{\partial \theta^2 \delta \Phi^A(\theta)} \,
\frac{\delta G}{\delta \bar{\Phi}_B(\theta)} \gamma_B^A +
(-1)^{\epsilon(F) \epsilon(G)} (F \leftrightarrow G) \Bigr\}.
\end{equation}

Notice, that $\{ F,G \}$ does {\em not} define a new superbracket since 
$\gamma_B^A=\delta_B^A \alpha(\phi^A)$ is a diagonal matrix. 
Taking into account the restriction 
(\ref{IV51}) the additional master equation (\ref{IV55}), at the lowest order 
of $\hbar$, simplifies according to
\begin{equation}
\label{IV57}
\int d^2 \theta \, \theta^2 \frac{\partial^2 \delta S_m}
{\partial \theta^2 \delta \Phi^A(\theta)} \, 
\Phi^B(\theta) \gamma_B^A + V_m S_m = 0.
\end{equation}
Obviously, the matrix $\gamma^B_A$ is uniquely determined by solving the
quantum master equations (\ref{IV47}) and (\ref{IV48}) at the lowest order
of $\hbar$, together with Eq. (\ref{IV57}). The matrix ${\bar\gamma}^B_A$, 
which enters in $V_m$, is fixed by the requirement (\ref{IIQ}). 
Equation (\ref{IV57}) expresses the conservation of the new ghost number of 
$S_m$ in the case $m \neq 0$, i.e., 
${\rm ngh}(S_m) = 0$. Thereby, we have formally ascribed also a new ghost 
number resp. Weyl weight to the mass parameter $m$, namely, according to the 
definition of $V_m$, ${\rm ngh}(m) = 1$ resp. $\alpha(m) = 1$. This already 
has been used in the definition of $V^a_m$, Eq. (\ref{IV44}).

Let us emphasize that the equation (\ref{IV55}) is quite formal since its 
right--hand side, for the same reasons as explained in the Introduction, is 
not well defined. Since, generally, the new ghost number is conserved only
in the classical limit, we restricted ourselves in (\ref{IV57}) to the lowest 
order approximation. In order to express the new ghost number conservation 
to higher orders -- which is, of course, only possible as long as the 
dilatation invariance in superspace is not broken by radiative corrections -- 
this requires a sensitive definition of the expression $\Delta_m W_m$ on 
the right--hand side of Eq. (\ref{IV55}). In order to obtain a corresponding 
local quantum operator equation, which is valid to all orders in perturbation 
theory, one can use the method described in Ref. $\cite{7}$.

Independently, by introducing a gauge the gauge--fixed quantum action 
(\ref{IV50}) breaks the new ghost number conservation. Namely, because of
\begin{equation*}
[ \Delta_m, \hat{T}_m(F) ] = \hbox{$\frac{1}{2}$} 
\epsilon_{ab} \{ \bar{\Delta}_m^b, [ \bar{\Delta}_m^a, 
[ \Delta_m, F ] + 2 F ] \} + (i/\hbar)^2 m^2 ( [ \Delta_m, F ] + 2 F ),
\end{equation*}
the action (\ref{IV50}) is only a solution of (\ref{IV57}) iff  
\begin{equation*}
[ \Delta_m, F ] W_m = - 2 F
\qquad
\Longrightarrow 
\qquad
\int d^2 \theta \, \theta^2 \frac{\partial^2 \delta F}
{\partial \theta^2 \delta \Phi^A(\theta)} \, 
\Phi^B(\theta) \gamma_B^A + V_m F = - 2 F,
\end{equation*}
where the second equation follows from the first one by taking into account
the condition (\ref{IV51}). On the other hand, the expression on the 
left--hand side (modulo the signum of $F$), can never be negative, 
since $F$ depends only on $\Phi^A$ which has positive Weyl weight,
\begin{equation*}
{\rm sgn}(F) \Bigr\{ \int d^2 \theta \, \theta^2 \frac{\partial^2 \delta F}
{\partial \theta^2 \delta \Phi^A(\theta)} \, 
\Phi^B(\theta) \gamma_B^A + V_m F \Bigr\} \geq 0.
\end{equation*}
This proves that the new ghost number conservation is broken through 
gauge fixing.


\section{Generating functionals and gauge (in)dependence}
\setcounter{equation}{0}


Next, we turn to the question of gauge (in)dependence of the generating 
functionals of Green's functions \cite{10,6}.

\smallskip
\noindent (A) {\it Sp(2)--covariant approach }
\\
In discussing this question it is convenient to study first the symmetry 
properties of the vacuum functional $Z(0)$ defined as
\begin{equation}
\label{V59}
Z(0) = \int d \Phi^A(\theta) \, d \bar{\Phi}_A(\theta) \,
\rho(\bar{\Phi}_A(\theta)) \exp\{ (i/ \hbar) ( W_{\rm ext} + S_X ) \}.
\end{equation}
Here, $\rho(\bar{\Phi}_A(\theta))$ is a density having the form of  
a $\delta$--functional,
\begin{equation}
\label{V60}
\rho(\bar{\Phi}_A(\theta)) = \delta \bigg(
\int d^2 \theta \, \bar{\Phi}_A(\theta) \bigg),
\end{equation}
and $S_X$ is given by 
\begin{equation}
\label{V61a}
S_X = \int d^2 \theta \, \bar{\Phi}_A(\theta) \Phi^A(\theta).
\end{equation}
The term $S_X$ can be cast into the (anti)BRST--invariant form
\begin{eqnarray*}
S_X = \hbox{$\frac{1}{2}$} \epsilon_{ab} \Bigr(
V^b ( V^a X - X U^a ) + ( V^a X - X U^a ) U^b \Bigr),
\qquad
X \equiv - \int d^2 \theta \, \theta^2 \bar{\Phi}_A(\theta) \Phi^A(\theta),
\end{eqnarray*} 
with $V^a \equiv V_+^a$ and $U^a \equiv U_+^a$, whose action on 
$\bar{\Phi}_A(\theta)$ and $\Phi^A(\theta)$ are defined in 
Eqs. (\ref{III31}) and (\ref{III36}), respectively, satisfying 
$\{ V^a, V^b \} = 0$ and $\{ U^a, U^b \} = 0$. Let us combine the action 
of $V^a$ and $U^a$ on an arbitrary functional $Y$ according to
\begin{equation*}
L^a Y \equiv V^a Y - (-1)^{\epsilon(Y)} Y U^a, 
\qquad
\{ L^a, L^b \} = 0,
\end{equation*}
then the operators $L^a$ are nilpotent and anticommuting. 

Inserting into expression (\ref{V59}) the relation (\ref{IV43}) and 
integrating by parts this gives
\begin{equation}
\label{V62}
Z(0) = \int d \Phi^A(\theta) \, d \bar{\Phi}_A(\theta) 
\rho(\bar{\Phi}_A(\theta))
\exp\{ (i/ \hbar) ( W + S_X + S_F ) \}
\end{equation}
with the following expression for $S_F$:
\begin{equation}
\label{V63}
S_F = - \int d^2 \theta \Bigl\{
\frac{\delta F}{\delta \Phi^A(\theta)} 
\frac{\partial^2 \Phi^A(\theta)}{\partial \theta^2} + 
\hbox{$\frac{1}{2}$} \epsilon_{ab} \int d^2 \bar{\theta} \,
\frac{\partial \Phi^A(\theta)}{\partial \theta_a}
\frac{\delta^2 F}{\delta \Phi^A(\theta) \delta \Phi^B(\bar{\theta)}}
\frac{\partial \Phi^B(\bar{\theta})}{\partial \bar{\theta}_b} \Bigr\}.
\end{equation}
This may be cast also into the (anti)BRST invariant form
\begin{equation*}
S_F = \hbox{$\frac{1}{2}$} \epsilon_{ab} F U^b U^a.
\end{equation*}
Then, by virtue of $L^a S_X = 0$ and $L^a S_F = 0$, it can be checked that 
the integrand of the vacuum functional (\ref{V62}) is invariant under the 
following 
global (anti)BRST transformations (thereby, one has to make use of 
Eq. (\ref{IV45})):
\begin{equation}
\label{V64}
\delta \Phi^A(\theta) = \Phi^A(\theta) U^a \mu_a,
\qquad
\delta \bar{\Phi}_A(\theta) = \mu_a V^a \bar{\Phi}_A(\theta) +
\mu_a ( W, \bar{\Phi}_A(\theta) )^a,
\end{equation}
where $\mu_a$, $\epsilon(\mu_a) = 1$, is a $Sp(2)$--doublet of constant
anticommuting parameters. Here, we have taken into account that the
density $\rho(\bar{\Phi}_A(\theta)) = \delta(\eta_A)$ is invariant under
the transformations (\ref{V64}). These transformations realize the
(anti)BRST symmetry in the superfield approach to quantum gauge theory.

The invariance of $Z(0)$ under the transformations (\ref{V64}) permits
to study the question whether $Z(0)$ is independent on the choice of the 
gauge. Indeed, let us change the gauge--fixing functional 
$F \rightarrow F + \delta F$. Then, the gauge--fixing term $S_F$ changes
according to 
\begin{equation}
\label{V65}
S_F \rightarrow S_{F + \delta F} = S_F + S_{\delta F},
\qquad
S_{\delta F} = \hbox{$\frac{1}{2}$} \epsilon_{ab} (\delta F) U^b U^a.
\end{equation}
Now, we perform in the vacuum functional (\ref{V62}) 
the transformations (\ref{V64}) and choose the parameters $\mu_a$ as follows,
\begin{equation*}
\mu_a = - (i/\hbar) \hbox{$\frac{1}{2}$} \epsilon_{ab} (\delta F) U^b.
\end{equation*}
Thereby we induce the factor ${\rm exp}(\mu_a U^a)$ in the integration 
measure. Combining its exponent with $S_F$ leads to
\begin{equation*}
S_F \rightarrow S_F + (\hbar/i) \mu_a U^a = S_F - 
\hbox{$\frac{1}{2}$} \epsilon_{ab} (\delta F) U^b U^a = S_F - S_{\delta F}.
\end{equation*}
By comparison with (\ref{V65}) this proves that the vacuum functional 
and, therefore, also the $S$--matrix is independent on the choice of the 
gauge.

\smallskip
\noindent (B) {\it osp(1,2)--covariant approach }
\\
In this approach the vacuum functional $Z_m(0)$, which depends on
the additional mass parameter $m$, is defined as
\begin{equation}
\label{V66}
Z_m(0) = \int d \Phi^A(\theta) \, d \bar{\Phi}_A(\theta) 
\rho(\bar{\Phi}_A(\theta)) \exp\{ (i/ \hbar) ( W_{m, {\rm ext}} + 
S_{m, X} ) \},
\end{equation}
with
\begin{equation}
S_{m, X} = S_X + m^2 \int d^2 \theta \, \theta^2 
\bar{\Phi}_A(\theta) \gamma^A_B \Phi^B(\theta),
\label{V61b}
\end{equation}
where $S_X$ again is given by Eq. (\ref{V61a}). The term $S_{m, X}$ can 
be rewritten as
\begin{equation*}
S_{m, X} = \hbox{$\frac{1}{2}$} \epsilon_{ab} \Big(
V_m^b ( V_m^a X - X U_m^a ) + ( V_m^a X - X U_m^a ) U_m^b \Big) + m^2 X,
\qquad
V_\alpha X + X U_\alpha = 0,
\end{equation*} 
with ($V_m^a \equiv V_+^a + \hbox{$\frac{1}{2}$} m^2 V_-^a, V_\alpha$) and 
($U_m^a \equiv U_+^a + \hbox{$\frac{1}{2}$} m^2 U_-^a, U_\alpha$) obeying
the following $osp(1,2)$--superalgebras
\begin{alignat}{2}
[ V_\alpha, V_\beta ] &= \epsilon_{\alpha\beta}^{~~~\!\gamma} V_\gamma, 
&\qquad\qquad
[ U_\alpha, U_\beta ] &= - \epsilon_{\alpha\beta}^{~~~\!\gamma} U_\gamma, 
\nonumber
\\
\label{V67}
[ V_\alpha, V_m^a ] &= V_m^b (\sigma_\alpha)_b^{~a}, 
&\qquad\qquad
[ U_\alpha, U_m^a ] &= - U_m^b (\sigma_\alpha)_b^{~a}, 
\\
\{ V_m^a, V_m^b \} &= - m^2 (\sigma^\alpha)^{ab} V_\alpha, 
&\qquad\qquad
\{ U_m^a, U_m^b \} &= m^2 (\sigma^\alpha)^{ab} U_\alpha, 
\nonumber
\end{alignat}
respectively; the action of ($V_\pm^a, V_\alpha$) and 
($U_\pm^a, U_\alpha$) on $\bar{\Phi}_A(\theta)$ and $\Phi^A(\theta)$ 
are defined by Eqs. (\ref{III32})--(\ref{III34}) and 
(\ref{III37})--(\ref{III39}), respectively.

Inserting into the expression (\ref{V66}) the relation (\ref{IV50}) 
and integrating by parts this yields
\begin{equation}
\label{V68}
Z_m(0) = \int d \Phi^A(\theta) \, d \bar{\Phi}_A(\theta) \,
\rho(\bar{\Phi}_A(\theta))
\exp\{ (i/ \hbar) ( W_m + S_{m, X} + S_{m, F} ) \},
\end{equation}
with
\begin{equation*}
S_{m, F} = S_F - \hbox{\large $\frac{1}{2}$} m^2 
\int d^2 \theta \, \theta^2 
\frac{\partial^2 \delta F}{\partial \theta^2 \delta \Phi^A(\theta)}
\gamma^A_B \Phi^B(\theta),
\end{equation*}
where $S_F$ is given by Eq.~(\ref{V63}). The gauge--fixing term $S_{m, F}$ 
can be rewritten as
\begin{equation*}
S_{m, F} = \hbox{$\frac{1}{2}$} \epsilon_{ab} F U_m^b U_m^a + m^2 F.
\end{equation*}

Let us now introduce the differential operators
\begin{equation*}
L_m^a Y \equiv V_m^a Y - (-1)^{\epsilon(Y)} Y U_m^a,
\qquad
L_\alpha Y \equiv V_\alpha Y + Y U_\alpha,
\end{equation*}
which, by virtue of the relations (\ref{V67}), satisfy the 
$osp(1,2)$--superalgebra
\begin{equation*}
[ L_\alpha, L_\beta ] = \epsilon_{\alpha\beta}^{~~~\!\gamma} L_\gamma,
\qquad
[ L_\alpha, L_m^a ] = L_m^b (\sigma_\alpha)_b^{~a},
\qquad
\{ L_m^a, L_m^b \} = - m^2 (\sigma^\alpha)^{ab} L_\alpha.
\end{equation*}
By using this algebra, after tedious but straightforward computations, 
one verifies the following relations:  
\begin{equation*}
L_m^c ( \hbox{$\frac{1}{2}$} \epsilon_{ab} L_m^b L_m^a + m^2 ) =
\hbox{$\frac{1}{2}$} m^2 (\sigma^\alpha)^c_{~d} L_m^d L_\alpha,
\qquad
[ L_\alpha, \hbox{$\frac{1}{2}$} \epsilon_{ab} L_m^b L_m^a + m^2 ] = 0.
\end{equation*}
Therefore, it holds $L_m^a S_{m, X} = 0$ and $L_\alpha S_{m, F} = 0$, since 
$X$ and $F$ are $Sp(2)$--invariant. Because $W_m$ exhibits the same 
$\eta$--dependence as $- S_{m, X}$, Eqs. (\ref{IV51}), (\ref{V61a}) and
(\ref{V61b}),\break $W_m + S_{m, X}$ is independent on $\eta_A$ and, hence, 
the integration over $\bar\Phi_A$ with the density 
$\rho(\bar{\Phi}_A(\theta)) = \delta(\eta_A)$ yields a constant factor which 
is equal to one.

We assert now that the integrand in (\ref{V68}) is invariant under the 
following global transformations (thereby, one has to make use of the 
Eqs. (\ref{IV47}) and (\ref{IV48}), respectively):
\begin{align}
\label{V69}
\delta \Phi^A(\theta) &= \Phi^A(\theta) U_m^a \mu_a,
\qquad
\delta \bar{\Phi}_A(\theta) = \mu_a V_m^a \bar{\Phi}_A(\theta) +
\mu_a ( W_m, \bar{\Phi}_A(\theta) )^a
\\
\label{V70}
\delta \Phi^A(\theta) &= \Phi^A(\theta) U_\alpha \mu^\alpha,
\qquad
\delta \bar{\Phi}_A(\theta) = \mu^\alpha V_\alpha \bar{\Phi}_A(\theta) +
\mu^\alpha \{ W_m, \bar{\Phi}_A(\theta) \}_\alpha,
\end{align}
where $\mu_a$, $\epsilon(\mu_a) = 1$, and $\mu^\alpha$, 
$\epsilon(\mu^\alpha) = 0$, are constant anticommuting resp.~commuting
parameters. Notice, that in the present case $\rho(\bar{\Phi}_A(\theta))$
is {\it not} invariant under the transformations (\ref{V69}). 
The transformations (\ref{V69}) and (\ref{V70}) realize the $m$--extended 
(anti)BRST-- and $Sp(2)$--symmetry, respectively. 

Next, we study the question whether the mass dependent terms in
$Z_m(0)$ violate the independence on the choice of the gauge. Proceeding
as in the previous case, by changing the gauge--fixing functional
$F \rightarrow F + \delta F$ the gauge--fixing term changes according to
\begin{equation}
\label{V71}
S_{m, F} \rightarrow S_{m, F + \delta F} = S_{m, F} + S_{m, \delta F},
\qquad
S_{m, \delta F} = \hbox{$\frac{1}{2}$} \epsilon_{ab} (\delta F) U_m^b U_m^a +
m^2 \delta F.
\end{equation}
Now, carring out in (\ref{V68}) the transformations (\ref{V69}), we choose
\begin{equation*}
\mu_a = - (i/\hbar) \hbox{$\frac{1}{2}$} \epsilon_{ab} (\delta F) U_m^b,
\end{equation*}
which leads to
\begin{equation*}
S_{m, F} \rightarrow S_{m, F} + (\hbar/i) \mu_a U_m^a = S_{m, F} - 
\hbox{$\frac{1}{2}$} \epsilon_{ab} (\delta F) U_m^b U_m^a = 
S_{m, F} - S_{m, \delta F} + m^2 \delta F.
\end{equation*}
By comparison with (\ref{V71}) we observe that the mass term $m^2 F$ violates 
the independence of $Z_m(0)$ on the choice of the gauge. One may try to 
compensate this undesired term $m^2 \delta F$ by means of an additional 
change of variables using the transformations (\ref{V70}). But this change
should not destroy the form of the action arrived at the previous stage.
However, such additional changes of variables lead to a Berezinian which
is equal to one because $\sigma_\alpha$ are traceless. Thus, the unwanted
term could never be compensated.


\section{Irreducible and first--stage reducible massive theories 
with closed algebra}
\setcounter{equation}{0}


In the preceeding Sections we gave a general framework of quantizing
massive general gauge theories by introducing on the space of superfields
and superantifields a set of differential operators which obey the
superalgebra $sl(1,2)$. Thereby, we extended our previous work \cite{6}
on $osp(1,2)$--covariant quantization where we already considered the
case of irreducible and first--stage reducible gauge theories with
closed algebra. In order to illustrate our present approach let us
study how the construction of these theories is extended now. (Thereby
we also simplify some of our former calculations.)

\smallskip
\noindent{\it (A) Generic form of the dependence on the antifields}\\ 
Our aim here is to construct a proper 
solution $S_m$ of the {\it classical} master equations
\begin{equation}
\label{VI72}
\hbox{$\frac{1}{2}$} ( S_m, S_m )^a + V_m^a S_m = 0,
\qquad
\hbox{$\frac{1}{2}$} \{ S_m, S_m \}_\alpha + V_\alpha S_m = 0,
\qquad
\hbox{$\frac{1}{2}$} \{ S_m, S_m \} + V_m S_m = 0,
\end{equation}
which are obtained from the quantum master equations (\ref{IV47}), 
(\ref{IV48}) 
and (\ref{IV55}) at the lowest order approximation of $\hbar$. 
Let us rewrite more explicit the brackets in Eqs. (\ref{VI72})
using their definitions, Eqs. (\ref{IV42}), (\ref{IV49}) and (\ref{IV56}), 
\begin{equation}
\label{VI73}
\frac{\delta S_m}{\delta \phi^A} 
\frac{\delta S_m}{\delta \phi_{A a}^*} + V_m^a S_m = 0,
\qquad
\frac{\delta S_m}{\delta \phi^A} 
\frac{\delta S_m}{\delta \eta_B} 
(\sigma_\alpha)_B^{~~\!A} + V_\alpha S_m = 0,
\qquad
\frac{\delta S_m}{\delta \phi^A} 
\frac{\delta S_m}{\delta \eta_B}
\gamma^B_A + V_m S_m = 0,
\end{equation}
with $V_m^a \equiv V_+^a + \hbox{$\frac{1}{2}$} m^2 V_-^a$ and 
$V_m \equiv V + m \partial/\partial m$, where the action of $V_\pm^a$,
$V_\alpha$ and $V$ on the antifields is given by 
(see Eqs. (\ref{II7}) and (\ref{II8})):
\begin{align*}
V_+^a &= \epsilon^{ab} \phi_{A b}^* \frac{\delta}{\delta \bar{\phi}_A} -
\eta_A \frac{\delta}{\delta \phi_{A a}^*},
\\
V_-^a &= \bar{\phi}_B \bigr(
(\sigma^\alpha)^a_{~b} (\sigma_\alpha)^B_{~~\!A} - 
\delta^a_b {\bar\gamma}^B_A \bigr)
\frac{\delta}{\delta \phi_{A b}^*} + \phi_{B b}^* \bigr(
(\sigma^\alpha)^{ab} (\sigma_\alpha)^B_{~~\!A} - 
\epsilon^{ab} ({\bar\gamma}^B_A + 2 \delta^B_A) \bigr)
\frac{\delta}{\delta \eta_A},
\\
V_\alpha &= \bar{\phi}_B (\sigma_\alpha)^B_{~~\!A}
\frac{\delta}{\delta \bar{\phi}_A} +
\bigr( \phi_{B b}^* (\sigma_\alpha)^B_{~~\!A} + 
\phi_{A a}^* (\sigma_\alpha)^a_{~b} \bigr)
\frac{\delta}{\delta \phi_{A b}^*} +
\eta_B (\sigma_\alpha)^B_{~~\!A}
\frac{\delta}{\delta \eta_A},
\\
V &= \bar{\phi}_B {\bar\gamma}^B_A
\frac{\delta}{\delta \bar{\phi}_A} +
\phi_{B b}^* ({\bar\gamma}^B_A + \delta^B_A)  
\frac{\delta}{\delta \phi_{A b}^*} +
\eta_B ({\bar\gamma}^B_A + 2 \delta^B_A)
\frac{\delta}{\delta \eta_A}.
\end{align*}
The symmetry properties (\ref{VI73}) of $S_m$ may be expressed also by the 
following equations: 
\begin{equation}
\label{VI74}
\mathbf{s}_m^a S_m = 0
\qquad
\mathbf{d}_\alpha S_m = 0,
\qquad
\mathbf{d}_m S_m = 0,
\end{equation}
with $\mathbf{s}_m^a \equiv \mathbf{s}_+^a + 
\hbox{$\frac{1}{2}$} m^2 \mathbf{s}_-^a$ and $\mathbf{d}_m \equiv 
\mathbf{d} + m \partial/\partial m$, where the operators 
$\mathbf{s}_\pm^a$, $\mathbf{d}_\alpha$ and $\mathbf{d}$ are 
required to fulfil the $sl(1,2)$--superalgebra:
\begin{alignat}{3}
[ \mathbf{d}, \mathbf{d}_\alpha ] &= 0,
&\qquad
[ \mathbf{d}, \mathbf{s}_+^a ] &= \mathbf{s}_+^a,
&\qquad
[ \mathbf{d}, \mathbf{s}_-^a ] &= - \mathbf{s}_-^a,
\nonumber
\\
\label{VI75}
[ \mathbf{d}_\alpha, \mathbf{d}_\beta ] &= 
\epsilon_{\alpha\beta}^{~~~\!\gamma} \mathbf{d}_\gamma, 
&\qquad 
[ \mathbf{d}_\alpha, \mathbf{s}_+^a ] &= 
\mathbf{s}_+^b (\sigma_\alpha)_b^{~a}, 
&\qquad 
[ \mathbf{d}_\alpha, \mathbf{s}_-^a ] &= 
\mathbf{s}_-^b (\sigma_\alpha)_b^{~a}, 
\\ 
\{ \mathbf{s}_+^a, \mathbf{s}_+^b \} &= 0,
&\qquad
\{ \mathbf{s}_-^a, \mathbf{s}_-^b \} &= 0,
&\qquad
\{ \mathbf{s}_+^a, \mathbf{s}_-^b \} &= 
- (\sigma^\alpha)^{ab} \mathbf{d}_\alpha - \epsilon^{ab} \mathbf{d}. 
\nonumber
\end{alignat}
Indeed, let us restrict our considerations to solutions $S_m$ being 
{\it linear} with respect to the antifields. 
Let us remark that proper solutions of the classical
master equations for theories with closed gauge algebra and vanishing 
new ghost number depends only linearly on the antifields \cite{13}.
Such solutions can be written in the form \cite{6}
\begin{equation}
\label{VI76}
S_m = S_{\rm cl} + (
\hbox{$\frac{1}{2}$} \epsilon_{ab} \mathbf{s}_m^b \mathbf{s}_m^a + m^2 ) X,
\end{equation}
where $X$ is assumed to be a $Sp(2)$--scalar (in fact the only one
we are able to build up linear in the antifields) and,
in accordance with the requirement (\ref{IIQ}), 
to have Weyl weight $\alpha(X) = \alpha(\bar{\phi}_A) + \alpha(\phi^A) = - 2$,
\begin{equation}
\label{VI77}
X = \bar{\phi}_A \phi^A
\qquad
{\rm with}
\qquad
\mathbf{d}_\alpha X = 0,
\qquad
\mathbf{d}_m X = - 2 X.
\end{equation}
Then, by making use of the $osp(1,2) \oplus u(1)$--superalgebra of these 
symmetry operators, 
\begin{alignat*}{3}
[ \mathbf{d}_m, \mathbf{d}_m ] &= 0,
&\qquad
[ \mathbf{d}_m, \mathbf{d}_\alpha ] &= 0,
&\qquad
[ \mathbf{d}_m, \mathbf{s}_m^a ] &= \mathbf{s}_m^a,
\\
[ \mathbf{d}_\alpha, \mathbf{d}_\beta ] &= 
\epsilon_{\alpha\beta}^{~~~\!\gamma} \mathbf{d}_\gamma, 
&\qquad 
[ \mathbf{d}_\alpha, \mathbf{s}_m^a ] &= 
\mathbf{s}_m^b (\sigma_\alpha)_b^{~a}, 
&\qquad 
\{ \mathbf{s}_m^a, \mathbf{s}_m^b \} &= 
- m^2 (\sigma^\alpha)^{ab} \mathbf{d}_\alpha, 
\end{alignat*}
one establishes the following relations: 
\begin{align*}
\mathbf{s}_m^c ( 
\hbox{$\frac{1}{2}$} \epsilon_{ab} \mathbf{s}_m^b \mathbf{s}_m^a + m^2 ) &=
\hbox{$\frac{1}{2}$} m^2 (\sigma^\alpha)^c_{~d} \mathbf{s}_m^d 
\mathbf{d}_\alpha, 
\\
\mathbf{d}_\alpha (
\hbox{$\frac{1}{2}$} \epsilon_{ab} \mathbf{s}_m^b \mathbf{s}_m^a + m^2 ) &= (
\hbox{$\frac{1}{2}$} \epsilon_{ab} \mathbf{s}_m^b \mathbf{s}_m^a + m^2 )
\mathbf{d}_\alpha,
\\
\mathbf{d}_m (
\hbox{$\frac{1}{2}$} \epsilon_{ab} \mathbf{s}_m^b \mathbf{s}_m^a + m^2 ) &= (
\hbox{$\frac{1}{2}$} \epsilon_{ab} \mathbf{s}_m^b \mathbf{s}_m^a + m^2 ) (
\mathbf{d}_m + 2 ).
\end{align*}
From these relations, by virtue of (\ref{VI77}), it follows that the ansatz 
(\ref{VI76}) for $S_m$ really obeys the symmetry requirements (\ref{VI74}). 
Thereby, it has to be taken into account that for the classical action 
$S_{\rm cl}(A)$ it holds $\mathbf{s}_m^a S_{\rm cl}(A) = 0$ as well as
$\mathbf{d}_\alpha S_{\rm cl}(A) = 0$ and $\mathbf{d}_m S_{\rm cl}(A) = 0$.

In order
to convince ourselves that the equations (\ref{VI74}) can be cast into the 
form (\ref{VI73}) let us decompose $\mathbf{s}_m^a$, $\mathbf{d}_\alpha$ and 
$\mathbf{d}_m$ into a component acting on the fields and another one acting 
on the antifields as follows:
\begin{equation}
\label{VI78}
\mathbf{s}_m^a = \left(\mathbf{s}_m^a \phi^A \right)
\frac{\delta_L}{\delta \phi^A} + V_m^a,
\qquad
\mathbf{d}_\alpha = \left(\mathbf{d}_\alpha \phi^A \right)
\frac{\delta_L}{\delta \phi^A} + V_\alpha,
\qquad
\mathbf{d}_m = \left(\mathbf{d}_m \phi^A \right)
\frac{\delta_L}{\delta \phi^A} + V_m.
\end{equation}
The assumptions (\ref{VI77}) are satisfied if the action of 
$\mathbf{d}_\alpha$ and $\mathbf{d}_m$ on $\phi^A$ is defined as 
\begin{equation*}
\mathbf{d}_\alpha \phi^A = \phi^B (\sigma_\alpha)_B^{~~\!A}
\qquad {\rm and} \qquad
\mathbf{d}_m \phi^A = \phi^B \gamma^A_B.
\end{equation*}

Then from (\ref{VI76}) one gets for $S_m$ the expression
\begin{equation*}
S_m = S_{\rm cl} + 
(\eta_A + \hbox{$\frac{1}{2}$} m^2 {\bar\gamma}^B_A \bar{\phi}_B ) \phi^A - 
(\mathbf{s}_m^a \phi^A) \phi_{A a}^* + \bar{\phi}_A ( 
\hbox{$\frac{1}{2}$} \epsilon_{ab} \mathbf{s}_m^b \mathbf{s}_m^a + m^2 ) 
\phi^A 
\end{equation*}
with ${\bar\gamma}^B_A = - \gamma^B_A - 2 \delta^B_A$. Now, performing in 
(\ref{VI74}) 
the replacements $\mathbf{s}_m^a \phi^A = - \delta_R S_m/\delta \phi_{A a}^*$, 
$\mathbf{d}_\alpha \phi^A = \delta S_m/\delta \eta_B
(\sigma_\alpha)_B^{~~\!A}$ and
$\mathbf{d}_m \phi^A = \delta S_m/\delta \eta_B \gamma_B^A$ it is
easily seen that both symmetry requirements, Eqs. (\ref{VI73}) and 
(\ref{VI74}), 
are equivalent to each other. Thus, we are left with the exercise to 
determine the action of the $sl(1,2)$--superalgebra (\ref{VI75}) on the
components of the fields $\phi^A$. Thereby, we restrict ourselves to the
cases of irreducible and first--stage reducible theories with closed
gauge algebra.

\smallskip
\noindent (B) {\it Explicit realization of sl(1,2) 
on the fields: Irreducible gauge theories}
\\
For irreducible theories with a closed algebra, because of
$M_{\alpha_0 \beta_0}^{ij} = 0$, the algebra of the generators, 
Eq. (\ref{II2}), reduces to 
\begin{equation}
\label{VI79}
R^i_{\alpha_0, j} R^j_{\beta_0} -
R^i_{\beta_0, j} R^j_{\alpha_0} = 
- R^i_{\gamma_0} F^{\gamma_0}_{\alpha_0 \beta_0},
\end{equation}
where for the sake of simplicity we assume throughout this and the succeeding
subsection that the $A^i$ are {\it bosonic} fields. This algebra defines
the structure tensors $F^{\gamma_0}_{\alpha_0 \beta_0}$. In general,
the restrictions imposed by the Jacobi identity lead to 
additional equations with new structure 
tensors. But in the simple case under consideration it leads 
only to the following relation among the tensors 
$F^{\gamma_0}_{\alpha_0 \beta_0}$ and the generators $R^i_{\alpha_0}$: 
\begin{equation}
\label{VI80}
F^{\delta_0}_{\eta_0 \alpha_0} F^{\eta_0}_{\beta_0 \gamma_0} -
R^i_{\alpha_0} F^{\delta_0}_{\beta_0 \gamma_0, i} +
\hbox{cyclic perm} (\alpha_0, \beta_0, \gamma_0) = 0.  
\end{equation}

In order to construct the proper solution $S_m = S_{\rm cl} + (
\hbox{$\frac{1}{2}$} \epsilon_{ab} \mathbf{s}_m^b \mathbf{s}_m^a + m^2) X$,
Eq. (\ref{VI76}), for $X$ one has to choose $X = \bar{A}_i A^i +
\bar{B}_{\alpha_0} B^{\alpha_0} + \bar{C}_{\alpha a} C^{\alpha_0 a}$. The
$sl(1,2)$--transformations of the antifields $\bar{A}_i$, $\bar{B}_{\alpha_0}$ 
and $\bar{C}_{\alpha_0 a}$ already has been given (see Appendix A). The 
corresponding {\em nonlinear} realization of the $sl(1,2)$ in terms of the 
fields $A^i$, $B^{\alpha_0}$ and $C^{\alpha_0 a}$ reads

\noindent{(1)~translations:}
\begin{align}
\mathbf{s}_+^a A^i &= R^i_{\alpha_0} C^{\alpha_0 a},
\nonumber
\\
\mathbf{s}_+^a C^{\alpha_0 b} &= \epsilon^{ab} B^{\alpha_0} -
F^{\alpha_0}_{\beta_0 \gamma_0} C^{\beta_0 a} C^{\gamma_0 b},
\label{VI81}
\\
\mathbf{s}_+^a B^{\alpha_0} &= \hbox{$\frac{1}{2}$} 
F^{\alpha_0}_{\beta_0 \gamma_0} B^{\beta_0} C^{\gamma_0 a} +
\hbox{$\frac{1}{12}$} \epsilon_{cd} (
F^{\alpha_0}_{\eta_0 \beta_0} F^{\eta_0}_{\gamma_0 \delta_0} +
2 R^i_{\beta_0} F^{\alpha_0}_{\gamma_0 \delta_0, i} )
C^{\gamma_0 a} C^{\delta_0 c} C^{\beta_0 d},
\nonumber
\\
\intertext{(2)~special conformal transformations:}
\mathbf{s}_-^a A^i &= 0,
\nonumber
\\
\mathbf{s}_-^a C^{\alpha_0 b} &= 0,
\label{VI82}
\\
\mathbf{s}_-^a B^{\alpha_0} &= - 2 C^{\alpha_0 a},
\nonumber
\\
\intertext{(3)~symplectic rotations:}
\mathbf{d}_\alpha A^i &= 0,
\nonumber
\\
\mathbf{d}_\alpha C^{\alpha_0 b} &= C^{\alpha_0 a} (\sigma_\alpha)_a^{~b},
\label{VI83}
\\
\mathbf{d}_\alpha B^{\alpha_0} &= 0,
\nonumber
\\
\intertext{and (4)~dilatations:}
\mathbf{d} A^i &= 0,
\nonumber
\\
\mathbf{d} C^{\alpha_0 b} &= C^{\alpha_0 b},
\label{VI84}
\\
\mathbf{d} B^{\alpha_0} &= 2 B^{\alpha_0}. 
\nonumber
\end{align}
By making use of Eqs. (\ref{VI79}) and (\ref{VI80}) it is a simple exercise to
prove that the transformations (\ref{VI81})--(\ref{VI84}) actually 
obey the $sl(1,2)$--superalgebra (\ref{VI75}). 
Let us remark that the nonlinearity of the translations, Eqs.~(\ref{VI81}),
 is due to the fact
that the components $\pi^{Aa}$ and $\lambda^a$ of the superfield
$\Phi^A(\theta)$ have been eliminated from the theory by integrating
them out in Eq. (\ref{V68}).

\smallskip
\noindent (C) {\it Explicit realization of sl(1,2) on the fields:
First--stage reducible gauge theories}
\\
Now let us consider first--stage reducible theories. In that case, due to 
the condition of first--stage reducibility,
\begin{equation}
\label{VI85}
R^i_{\alpha_0} Z^{\alpha_0}_{\alpha_1} = 0,
\end{equation}
there are independent zero--modes $Z^{\alpha_0}_{\alpha_1}$ of the generators 
$R^i_{\alpha_0}$. Their presence does not modify the gauge algebra  
\begin{equation}
\label{VI86}
R^i_{\alpha_0, j} R^j_{\beta_0} -
R^i_{\beta_0, j} R^j_{\alpha_0} = 
- R^i_{\gamma_0} F^{\gamma_0}_{\alpha_0 \beta_0},
\end{equation}
but it influences the solutions of the Jacobi identity 
which appears from the relation
\begin{equation}
\label{VI87}
R^j_{\delta_0} \bigr(
F^{\delta_0}_{\eta_0 \alpha_0} F^{\eta_0}_{\beta_0 \gamma_0} -
R^i_{\alpha_0} F^{\delta_0}_{\beta_0 \gamma_0, i} +
\hbox{cyclic perm} (\alpha_0, \beta_0, \gamma_0) \bigr) = 0.  
\end{equation}
In addition, new equations and structure tensors occure. One of
these gauge structure relations is the reducibility condition 
(\ref{VI85}) itself. In order to derive the others we proceed as
follows:
 
First, let us cast the Jacobi identity (\ref{VI87}) into a more practical 
form. Owing to (\ref{VI85}) the expression in parenthesis must be 
proportional to the zero--modes $Z^{\delta_0}_{\alpha_1}$,  
\begin{equation}
\label{VI88}
F^{\delta_0}_{\eta_0 \alpha_0} F^{\eta_0}_{\beta_0 \gamma_0} -
R^i_{\alpha_0} F^{\delta_0}_{\beta_0 \gamma_0, i} +
\hbox{cyclic perm} (\alpha_0, \beta_0, \gamma_0) =
3 Z^{\delta_0}_{\alpha_1} H^{\alpha_1}_{\alpha_0 \beta_0 \gamma_0},
\end{equation}
where $H^{\alpha_1}_{\alpha_0 \beta_0 \gamma_0}(A)$ are new structure
tensors being totally antisymmetric with respect to the indices 
$\alpha_0$, $\beta_0$, $\gamma_0$ and depending, in general, on the gauge 
fields $A^i$. 

Next, we derive an expression for the combination 
$Z^{\alpha_0}_{\alpha_1, j} R^j_{\beta_0}$. Multiplying (\ref{VI86}) by 
$Z^{\alpha_0}_{\alpha_1}$ and using the relation 
$R^i_{\alpha_0, j} Z^{\alpha_0}_{\alpha_1} =
- R^i_{\alpha_0} Z^{\alpha_0}_{\alpha_1, j}$, which follows from (\ref{VI85}), 
we obtain
\begin{equation*}
R^i_{\alpha_0} (
Z^{\alpha_0}_{\alpha_1, j} R^j_{\beta_0} +
F^{\alpha_0}_{\beta_0 \gamma_0} Z^{\gamma_0}_{\alpha_1} ) = 0.
\end{equation*}
Again, this may be solved by introducing additional structure tensors 
$G^{\gamma_1}_{\beta_0 \alpha_1}(A)$ 
\begin{equation}
\label{VI89}
Z^{\alpha_0}_{\alpha_1, j} R^j_{\beta_0} +
F^{\alpha_0}_{\beta_0 \gamma_0} Z^{\gamma_0}_{\alpha_1} = 
- Z^{\alpha_0}_{\gamma_1} G^{\gamma_1}_{\beta_0 \alpha_1},
\end{equation}
thus defining a new structure equation for first--stage reducible
theories. Multiplying this equation by $Z^{\beta_0}_{\beta_1}$ and 
taking into account (\ref{VI85}),
\begin{equation*}
F^{\alpha_0}_{\beta_0 \gamma_0}
Z^{\gamma_0}_{\alpha_1} Z^{\beta_0}_{\beta_1} =
- Z^{\alpha_0}_{\gamma_1} Z^{\beta_0}_{\beta_1} 
G^{\gamma_1}_{\beta_0 \alpha_1}, 
\end{equation*}
we obtain the useful equality
\begin{equation}
\label{VI90}
Z^{\alpha_0}_{\beta_1} G^{\gamma_1}_{\alpha_0 \alpha_1} = 
- Z^{\alpha_0}_{\alpha_1} G^{\gamma_1}_{\alpha_0 \beta_1}.
\end{equation}

Moreover, we are able to establish two further gauge structure relations 
for the first--stage reducible case showing that 
$H^{\alpha_1}_{\alpha_0 \beta_0 \gamma_0}$ and 
$G^{\alpha_1}_{\alpha_0 \beta_1}$ are not independent of each other. The 
first one reads
\begin{equation}
\label{VI91}
\bigr(
G^{\alpha_1}_{\beta_0 \gamma_1} G^{\gamma_1}_{\gamma_0 \beta_1} +
R^i_{\beta_0} G^{\alpha_1}_{\gamma_0 \beta_1, i} +
\hbox{antisym}(\beta_0 \leftrightarrow \gamma_0) \bigr) +
G^{\alpha_1}_{\alpha_0 \beta_1} F^{\alpha_0}_{\beta_0 \gamma_0} +
3 Z^{\alpha_0}_{\beta_1} H^{\alpha_1}_{\alpha_0 \beta_0 \gamma_0} = 0.
\end{equation}
In order to verify this relation we multiply the Jacobi identity (\ref{VI88})
with $Z^{\alpha_0}_{\beta_1}$. By virtue of 
$R^i_{\alpha_0} Z^{\alpha_0}_{\beta_1} = 0$ this yields 
\begin{align*}
& ( F^{\delta_0}_{\eta_0 \alpha_0} Z^{\alpha_0}_{\beta_1} ) 
F^{\eta_0}_{\beta_0 \gamma_0} +
F^{\delta_0}_{\eta_0 \beta_0}  
( F^{\eta_0}_{\gamma_0 \alpha_0} Z^{\alpha_0}_{\beta_1} ) -
F^{\delta_0}_{\eta_0 \gamma_0}  
( F^{\eta_0}_{\beta_0 \alpha_0} Z^{\alpha_0}_{\beta_1} )
\\
& ~ - R^i_{\beta_0} (
F^{\delta_0}_{\gamma_0 \alpha_0, i} Z^{\alpha_0}_{\beta_1} ) +
R^i_{\gamma_0} (
F^{\delta_0}_{\beta_0 \alpha_0, i} Z^{\alpha_0}_{\beta_1} ) -
Z^{\delta_0}_{\alpha_1} ( 
3 Z^{\alpha_0}_{\beta_1} H^{\alpha_1}_{\alpha_0 \beta_0 \gamma_0} ) = 0.
\end{align*}
After replacing all terms of the form  
$F^{\delta_0}_{\eta_0 \alpha_0} Z^{\alpha_0}_{\beta_1}$ according to 
(\ref{VI89}) this gives
\begin{align*}
& Z^{\delta_0}_{\beta_1, i} 
( R^i_{\alpha_0} F^{\alpha_0}_{\beta_0 \gamma_0} ) +
Z^{\delta_0}_{\alpha_1} ( 
G^{\alpha_1}_{\alpha_0 \beta_1} F^{\alpha_0}_{\beta_0 \gamma_0} +
3 Z^{\alpha_0}_{\beta_1} H^{\alpha_1}_{\alpha_0 \beta_0 \gamma_0} )
\\
& ~ + \bigr\{ R^i_{\beta_0} (
F^{\delta_0}_{\gamma_0 \alpha_0, i} Z^{\alpha_0}_{\beta_1} -
F^{\delta_0}_{\alpha_0 \gamma_0} Z^{\alpha_0}_{\beta_1, i} ) -
( F^{\delta_0}_{\alpha_0 \gamma_0} Z^{\alpha_0}_{\alpha_1} )
G^{\alpha_1}_{\beta_0 \beta_1} +
\hbox{antisym}(\beta_0 \leftrightarrow \gamma_0) \bigr\} = 0,
\end{align*}
and, using the same relation once more, 
\begin{align*}
& Z^{\delta_0}_{\beta_1, i} 
( R^i_{\alpha_0} F^{\alpha_0}_{\beta_0 \gamma_0} ) + 
Z^{\delta_0}_{\alpha_1} ( 
G^{\alpha_1}_{\alpha_0 \beta_1} F^{\alpha_0}_{\beta_0 \gamma_0} +
3 Z^{\alpha_0}_{\beta_1} H^{\alpha_1}_{\alpha_0 \beta_0 \gamma_0} )
\\
& ~ + \bigr\{ R^i_{\beta_0} \bigr(
( F^{\delta_0}_{\gamma_0 \alpha_0} Z^{\alpha_0}_{\beta_1} )_{,i} +
Z^{\delta_0}_{\alpha_1, i} G^{\alpha_1}_{\gamma_0 \beta_1} \bigr) +
Z^{\delta_0}_{\alpha_1}  
G^{\alpha_1}_{\beta_0 \gamma_1} G^{\gamma_1}_{\gamma_0 \beta_1} +
\hbox{antisym}(\beta_0 \leftrightarrow \gamma_0) \bigr\} = 0.
\end{align*}
Here, the expression in the curly bracket can be rewritten as
\begin{align*}
& Z^{\delta_0}_{\beta_1, i} 
( R^i_{\alpha_0} F^{\alpha_0}_{\beta_0 \gamma_0} ) +
Z^{\delta_0}_{\alpha_1} ( 
G^{\alpha_1}_{\alpha_0 \beta_1} F^{\alpha_0}_{\beta_0 \gamma_0} +
3 Z^{\alpha_0}_{\beta_1} H^{\alpha_1}_{\alpha_0 \beta_0 \gamma_0} )
\\
& ~ + \bigr\{ R^i_{\beta_0} 
( F^{\delta_0}_{\gamma_0 \alpha_0} Z^{\alpha_0}_{\beta_1} +
Z^{\delta_0}_{\alpha_1} G^{\alpha_1}_{\gamma_0 \beta_1} )_{,i} +
Z^{\delta_0}_{\alpha_1} ( 
G^{\alpha_1}_{\gamma_1 \beta_0} G^{\gamma_1}_{\gamma_0 \beta_1} +
R^i_{\gamma_0} G^{\alpha_1}_{\beta_0 \beta_1, i} ) +
\hbox{antisym}(\beta_0 \leftrightarrow \gamma_0) \bigr\} = 0
\end{align*}
and furthermore, once again using relation (\ref{VI89}),  
\begin{align}
& Z^{\delta_0}_{\beta_1, i} 
( R^i_{\alpha_0} F^{\alpha_0}_{\beta_0 \gamma_0} ) +
Z^{\delta_0}_{\alpha_1} ( 
G^{\alpha_1}_{\alpha_0 \beta_1} F^{\alpha_0}_{\beta_0 \gamma_0} +
3 Z^{\alpha_0}_{\beta_1} H^{\alpha_1}_{\alpha_0 \beta_0 \gamma_0} )
\nonumber\\
& ~ - \bigr\{ R^i_{\beta_0} 
( Z^{\delta_0}_{\beta_1, j} R^j_{\gamma_0} )_{,i} -
Z^{\delta_0}_{\alpha_1} ( 
G^{\alpha_1}_{\beta_0 \gamma_1} G^{\gamma_1}_{\gamma_0 \beta_1} +
R^i_{\gamma_0} G^{\alpha_1}_{\beta_0 \beta_1, i} ) +
\hbox{antisym}(\beta_0 \leftrightarrow \gamma_0) \bigr\} = 0.
\end{align}
This equation, since the algebra (\ref{VI86}) is closed,
\begin{equation*}
Z^{\delta_0}_{\beta_1, i} 
( R^i_{\alpha_0} F^{\alpha_0}_{\beta_0 \gamma_0} ) =
Z^{\delta_0}_{\beta_1, i} 
( R^j_{\beta_0} R^i_{\gamma_0, j} - R^j_{\gamma_0} R^i_{\beta_0, j} ) =
R^i_{\beta_0} ( Z^{\delta_0}_{\beta_1, j} R^j_{\gamma_0} )_{,i} -
R^i_{\gamma_0} ( Z^{\delta_0}_{\beta_1, j} R^j_{\beta_0} )_{,i},
\end{equation*}
leads immediately to the gauge structure relation (\ref{VI91}).

The second gauge structure relation, which can also be derived by means of 
the Jacobi identity, is given by
\begin{align}
\label{VI92}
\bigr(&
H^{\alpha_1}_{\eta_0 \alpha_0 \beta_0} F^{\eta_0}_{\gamma_0 \delta_0} -
H^{\alpha_1}_{\eta_0 \delta_0 \alpha_0} F^{\eta_0}_{\beta_0 \gamma_0} + 
\hbox{cyclic perm} (\alpha_0, \beta_0, \gamma_0) \bigr)
\nonumber
\\
& + \bigr\{ 
R^i_{\delta_0} H^{\alpha_1}_{\alpha_0 \beta_0 \gamma_0, i} -
G^{\alpha_1}_{\delta_0 \beta_1} H^{\beta_1}_{\alpha_0 \beta_0 \gamma_0} +
{\rm antisym}\bigr(
\delta_0 \leftrightarrow (\alpha_0, \beta_0, \gamma_0) \bigr) \bigr\} = 0,
\end{align}
where the left--hand side is a totally antisymmetric expression with respect 
to $(\alpha_0, \beta_0, \gamma_0, \delta_0)$.

In order to prove that this relation is satisfied we consider the following 
identity:
\begin{align*}
\bigr\{&
\bigr( ( Z^{\lambda_0}_{\alpha_1} H^{\alpha_1}_{\eta_0 \alpha_0 \beta_0} )
F^{\eta_0}_{\gamma_0 \delta_0} + 
\hbox{cyclic perm} (\alpha_0, \beta_0, \gamma_0) \bigr)
\\
& + 2 R^i_{\delta_0}
( Z^{\lambda_0}_{\alpha_1} H^{\alpha_1}_{\alpha_0 \beta_0 \gamma_0} )_{,i} +
2 F^{\lambda_0}_{\delta_0 \eta_0} 
( Z^{\eta_0}_{\alpha_1} H^{\alpha_1}_{\alpha_0 \beta_0 \gamma_0} )
\bigr\}  + {\rm antisym}\bigr(
\delta_0 \leftrightarrow (\alpha_0, \beta_0, \gamma_0) \bigr) \equiv 0,
\end{align*}
which can be verified by a direct calculation replacing the terms
$Z^{\lambda_0}_{\alpha_1} H^{\alpha_1}_{\alpha_0 \beta_0 \gamma_0}$ by the 
help of the Jacobi identity (\ref{VI88}). Taking into account (\ref{VI89}) 
one obtains the equation
\begin{align*}
Z^{\lambda_0}_{\alpha_1} \bigr\{& \bigr(
H^{\alpha_1}_{\eta_0 \alpha_0 \beta_0} F^{\eta_0}_{\gamma_0 \delta_0} + 
\hbox{cyclic perm} (\alpha_0, \beta_0, \gamma_0) \bigr)
\\
& + 2 R^i_{\delta_0} H^{\alpha_1}_{\alpha_0 \beta_0 \gamma_0, i} -
2 G^{\alpha_1}_{\beta_1 \delta_0} H^{\beta_1}_{\alpha_0 \beta_0 \gamma_0} 
\bigr\} + {\rm antisym}\bigr(
\delta_0 \leftrightarrow (\alpha_0, \beta_0, \gamma_0) \bigr) = 0.
\end{align*}
After factoring out the zero--modes $Z^{\lambda_0}_{\alpha_1}$ and using
the identity
\begin{align*}
\bigr(&
H^{\alpha_1}_{\eta_0 \alpha_0 \beta_0} F^{\eta_0}_{\gamma_0 \delta_0} + 
\hbox{cyclic perm} (\alpha_0, \beta_0, \gamma_0) \bigr) + 
{\rm antisym}\bigr(
\delta_0 \leftrightarrow (\alpha_0, \beta_0, \gamma_0) \bigr)
\\
& \equiv 2 \bigr(
H^{\alpha_1}_{\eta_0 \alpha_0 \beta_0} F^{\eta_0}_{\gamma_0 \delta_0} - 
H^{\alpha_1}_{\eta_0 \delta_0 \alpha_0} F^{\eta_0}_{\beta_0 \gamma_0} + 
\hbox{cyclic perm} (\alpha_0, \beta_0, \gamma_0) \bigr),
\end{align*}
this equation acquires the form (\ref{VI92}). The relations
(\ref{VI84})--(\ref{VI92}) are the key equations in order to derive the
$sl(1,2)$--transformations of the fields for the first--stage reducible case. 

In order to construct the proper solution $S_m = S_{\rm cl} + 
(\hbox{$\frac{1}{2}$} \epsilon_{ab} \mathbf{s}_m^b \mathbf{s}_m^a + m^2) X$
in that case one has to choose 
$X = {\bar A}_i A^i + {\bar B}_{\alpha_0} B^{\alpha_0} +
{\bar B}_{\alpha_1 a} B^{\alpha_1 a} + {\bar C}_{\alpha_0 a} C^{\alpha_0 a} +
{\bar C}_{\alpha_1 ab} C^{\alpha_1 ab}$. A realization of the 
$sl(1,2)$--transformations of the antifields 
${\bar A}_i$, ${\bar B}_{\alpha_0}$, ${\bar B}_{\alpha_1 a}$, 
${\bar C}_{\alpha_0 a}$ and ${\bar C}_{\alpha_1 ab}$ already has been given 
(see Appendix A). 
The corresponding nonlinear realization of the 
$sl(1,2)$ in terms of the fields $A^i$, 
$B^{\alpha_0}$, $B^{\alpha_1 a}$, $C^{\alpha_0 a}$ and $C^{\alpha_1 ab}$ 
are the following

\noindent{(1)~translations:}
\begin{align}
\mathbf{s}_+^a A^i &= R^i_{\alpha_0} C^{\alpha_0 a},
\nonumber
\\
\mathbf{s}_+^a C^{\alpha_0 b} &= Z^{\alpha_0}_{\alpha_1} C^{\alpha_1 ab} +
\epsilon^{ab} B^{\alpha_0} -
F^{\alpha_0}_{\beta_0 \gamma_0} C^{\beta_0 a} C^{\gamma_0 b},
\nonumber
\\
\mathbf{s}_+^a B^{\alpha_0} &= Z^{\alpha_0}_{\alpha_1} B^{\alpha_1 a} +
\hbox{$\frac{1}{2}$} F^{\alpha_0}_{\beta_0 \gamma_0} (
B^{\beta_0} C^{\gamma_0 a} -
\epsilon_{cd} Z^{\beta_0}_{\alpha_1} C^{\alpha_1 ac} C^{\gamma_0 d} )
\nonumber
\\
&\quad~ + \hbox{$\frac{1}{12}$} \epsilon_{cd} (
F^{\alpha_0}_{\eta_0 \beta_0} F^{\eta_0}_{\gamma_0 \delta_0} +
2 R^i_{\beta_0} F^{\alpha_0}_{\gamma_0 \delta_0, i} )
C^{\gamma_0 a} C^{\delta_0 c} C^{\beta_0 d},
\label{VI93}
\\
\mathbf{s}_+^a C^{\alpha_1 bc} &= - \epsilon^{ac} B^{\alpha_1 b} -
\epsilon^{ab} B^{\alpha_1 c} +
G^{\alpha_1}_{\alpha_0 \beta_1} C^{\alpha_0 a} C^{\beta_1 bc} -
\hbox{$\frac{1}{2}$} H^{\alpha_1}_{\alpha_0 \beta_0 \gamma_0}
C^{\alpha_0 a} C^{\beta_0 b} C^{\gamma_0 c},
\nonumber
\\
\mathbf{s}_+^a B^{\alpha_1 b} &= G^{\alpha_1}_{\alpha_0 \beta_1} (
C^{\alpha_0 a} B^{\beta_1 b} - 
\hbox{$\frac{1}{2}$} \epsilon_{cd} 
Z^{\alpha_0}_{\gamma_1} C^{\gamma_1 ac} C^{\beta_1 bd} ) - 
\hbox{$\frac{1}{2}$} H^{\alpha_1}_{\alpha_0 \beta_0 \gamma_0} 
B^{\alpha_0} C^{\beta_0 a} C^{\gamma_0 c}
\nonumber
\\
& \quad ~ + \hbox{$\frac{1}{4}$} \epsilon_{cd}
H^{\alpha_1}_{\alpha_0 \beta_0 \gamma_0} Z^{\alpha_0}_{\beta_1} (
3 C^{\beta_0 a} C^{\beta_1 bc} C^{\gamma_0 d} +
C^{\beta_0 b} C^{\beta_1 ac} C^{\gamma_0 d} )
\nonumber
\\
& \quad~ + \hbox{$\frac{1}{8}$} \epsilon_{cd} (
G^{\alpha_1}_{\delta_0 \beta_1} H^{\beta_1}_{\alpha_0 \beta_0 \gamma_0} -
R^i_{\delta_0} H^{\alpha_1}_{\alpha_0 \beta_0 \gamma_0, i} )
C^{\gamma_0 a} C^{\beta_0 b} C^{\alpha_0 c} C^{\delta_0 d}
\nonumber
\\
& \quad~ - \hbox{$\frac{1}{16}$} \epsilon_{cd} 
H^{\alpha_1}_{\eta_0 \alpha_0 \beta_0} F^{\eta_0}_{\gamma_0 \delta_0} (
C^{\gamma_0 a} C^{\beta_0 b} + C^{\gamma_0 b} C^{\beta_0 a} )
C^{\alpha_0 c} C^{\delta_0 d},  
\nonumber
\\
\intertext{(2)~special conformal transformations:}
\mathbf{s}_-^a A^i &= 0,
\nonumber
\\
\mathbf{s}_-^a C^{\alpha_0 b} &= 0,
\nonumber
\\
\mathbf{s}_-^a B^{\alpha_0} &= - 2 C^{\alpha_0 a},
\label{VI94}
\\
\mathbf{s}_-^a C^{\alpha_1 bc} &= 0,
\nonumber
\\
\mathbf{s}_-^a B^{\alpha_1 b} &= 2 C^{\alpha_1 ab},
\nonumber
\\
\intertext{(3)~symplectic rotations:}
\mathbf{d}_\alpha A^i &= 0,
\nonumber
\\
\mathbf{d}_\alpha C^{\alpha_0 b} &= C^{\alpha_0 a} (\sigma_\alpha)_a^{~b},
\nonumber
\\
\mathbf{d}_\alpha B^{\alpha_0} &= 0,
\label{VI95}
\\
\mathbf{d}_\alpha C^{\alpha_1 bc} &= C^{\alpha_1 ac} (\sigma_\alpha)_a^{~b} +
C^{\alpha_1 ba} (\sigma_\alpha)_a^{~c},
\nonumber
\\
\mathbf{d}_\alpha B^{\alpha_1 b} &= B^{\alpha_1 a} (\sigma_\alpha)_a^{~b},
\nonumber
\\
\intertext{and (4)~dilatations:}
\mathbf{d} A^i &= 0,
\nonumber
\\
\mathbf{d} C^{\alpha_0 b} &= C^{\alpha_0 b},
\nonumber
\\
\mathbf{d} B^{\alpha_0} &= 2 B^{\alpha_0}, 
\label{VI96}
\\
\mathbf{d} C^{\alpha_1 bc} &= 2 C^{\alpha_1 bc},
\nonumber
\\
\mathbf{d} B^{\alpha_1 b} &= 3 B^{\alpha_1 b}.
\nonumber
\end{align}
By making use of Eqs. (\ref{VI84})--(\ref{VI92}) after somewhat involved and
tedious algebraic mani\-pulations it can be proven that the transformations
(\ref{VI93})--(\ref{VI96}) really obey the $sl(1,2)$--superalgebra 
(\ref{VI75}). For some details of this work we refer to Ref.~\cite{6} 
where similar calculations were performed for the $osp(1,2)$--superalgebra.

Continuing in the same way, analogous considerations can be made for 
higher stage reducible theories. But then, more and more new gauge 
structure tensors with increasing numbers of indices and additional gauge 
structure relations appear which makes a study of these theories quite 
complicated. 


\section{\bf Concluding remarks} 


In this paper we have revealed the geometrical content of the 
$osp(1,2)$--covariant Lagrangian quantization of general massive gauge 
theories. A natural geometric formulation of that quantization procedure
is obtained by considering $osp(1,2)$ as subsuperalgebra of
$sl(1,2)$, which is considered as the algebra of generators of conformal 
transformations in two anticommuting dimensions. It is shown that proper
solutions of the classical master equations can be constructed being
invariant under $osp(1,2) \oplus u(1)$. The $m$--dependent extended BRST
symmetry is realized in superspace as translations combined with 
$m$--dependent special conformal transformations. The $sl(2) \oplus u(1)$
symmetry is realized in superspace as symplectic rotations and dilatations,
respectively. By the choice of a gauge the $sl(2) \oplus u(1)$ symmetry 
is broken down to $sl(2) \sim sp(2)$.
In principle, by formal manipulations it is also possible to construct proper 
solutions of the corresponding quantum master equations. However, in doing 
so a serious problem is to provide a sensible definition of the various
$\Delta$--operators of the quantum master equations, 
which do not make sense when applied to local functionals. 
In this paper we have not adressed such problems
and related questions as the use of explicit regularizations and
renormalizations schemes and the discussion of the role of anomalies. 
 
\bigskip
\bigskip
\noindent
{\large\bf Acknowledgement}\\
The authors would like to thank P.M. Lavrov for valuable discussions
concerning various aspects of the superfield quantization of general
gauge theories.

\bigskip
\bigskip

\begin{appendix}


\section{Componentwise notation of the $sl(1,2)$ transformations
of the antifields}


In componentwise notation the linear transformations (\ref{II7}) generated by 
$V_+^a$ and $V_-^a$ read as follows:  
\begin{align*}
V_+^a \bar{A}_i &= \epsilon^{ab} A_{i b}^*,
\\
V_+^a A^*_{i b} &= - \delta^a_b D_i,
\\
V_+^a D_i &= 0,
\\
V_+^a \bar{B}_{\alpha_s|a_1 \cdots a_s} &= \epsilon^{ab}
B_{\alpha_s b|a_1 \cdots a_s}^*,
\\
V_+^a B_{\alpha_s b|a_1 \cdots a_s}^* &= - \delta^a_b 
E_{\alpha_s|a_1 \cdots a_s},
\\
V_+^a E_{\alpha_s|a_1 \cdots a_s} &= 0,
\\
V_+^a \bar{C}_{\alpha_s|a_0 \cdots a_s} &= \epsilon^{ab}
C_{\alpha_s b|a_0 \cdots a_s}^*,
\\
V_+^a C_{\alpha_s b|a_0 \cdots a_s}^* &= - \delta^a_b 
F_{\alpha_s|a_0 \cdots a_s},
\\
V_+^a F_{\alpha_s|a_0 \cdots a_s} &= 0
\\
\intertext{and}
V_-^a \bar{A}_i &= 0,
\\
V_-^a A_{i b}^* &= 2 \delta^a_b \bar{A}_i,
\\
V_-^a D_i &= 0,
\\
V_-^a \bar{B}_{\alpha_s|a_1 \cdots a_s} &= 0,
\\
V_-^a B_{\alpha_s b|a_1 \cdots a_s}^* &=
2 \delta^a_b \Bigr(
\bar{B}_{\alpha_s|a_1 \cdots a_s} +
\sum_{r = 1}^s \delta^a_{a_r}
\bar{B}_{\alpha_s|a_1 \cdots a_{r - 1} b a_{r + 1} \cdots a_s} \Bigr),
\\
V_-^a E_{\alpha_s|a_1 \cdots a_s} &= 2 \epsilon^{ab} \Bigr(
(s + 2) B_{\alpha_s b|a_1 \cdots a_s} -
\sum_{r = 1}^s B_{\alpha_s a_r|a_1 \cdots a_{r - 1} b a_{r + 1} \cdots a_s}^*
\Bigr),
\\
V_-^a \bar{C}_{\alpha_s|a_0 \cdots a_s} &= 0,
\\
V_-^a C_{\alpha_s b|a_0 \cdots a_s}^* &= 
2 \delta^a_b \Bigr(
\bar{C}_{\alpha_s|a_0 \cdots a_s} +
\sum_{r = 0}^s \delta^a_{a_r}
\bar{C}_{\alpha_s|a_0 \cdots a_{r - 1} b a_{r + 1} \cdots a_s} \Bigr),
\\
V_-^a F_{\alpha_s|a_0 \cdots a_s} &= 2 \epsilon^{ab} \Bigr( 
(s + 1) C_{\alpha_s b|a_0 \cdots a_s} -
\sum_{r = 0}^s C_{\alpha_s a_r|a_0 \cdots a_{r - 1} b
a_{r + 1} \cdots a_s}^* \Bigr),
\end{align*}
where the definitions (\ref{II9}) and (\ref{II10}) have to be taken into
account. For the transformations (\ref{II8}) generated by $V_\alpha$ and $V$ 
one gets:
\begin{align*}
V_\alpha \bar{A}_i &= 0,
\\
V_\alpha A_{i b}^* &= A_{i a}^* (\sigma_\alpha)^a_{~b},
\\
V_\alpha D_i &= 0, 
\\
V_\alpha \bar{B}_{\alpha_s|a_1 \cdots a_s} &= \sum_{r = 1}^s
\bar{B}_{\alpha_s|a_1 \cdots a_{r - 1} a a_{r + 1} \cdots a_s} 
(\sigma_\alpha)^a_{~a_r},
\\
V_\alpha B_{\alpha_s b|a_1 \cdots a_s}^* &= 
B_{\alpha_s a|a_1 \cdots a_s}^* (\sigma_\alpha)^a_{~b} + \sum_{r = 1}^s
B_{\alpha_s b|a_1 \cdots a_{r - 1} a a_{r + 1} \cdots a_s}^* 
(\sigma_\alpha)^a_{~a_r},
\\
V_\alpha E_{\alpha_s|a_1 \cdots a_s} &= \sum_{r = 1}^s
E_{\alpha_s|a_1 \cdots a_{r - 1} a a_{r + 1} \cdots a_s} 
(\sigma_\alpha)^a_{~a_r},
\\
V_\alpha \bar{C}_{\alpha_s|a_0 \cdots a_s} &= \sum_{r = 0}^s
\bar{C}_{\alpha_s|a_0 \cdots a_{r - 1} a a_{r + 1} \cdots a_s} 
(\sigma_\alpha)^a_{~a_r},
\\
V_\alpha C_{\alpha_s b|a_0 \cdots a_s}^* &= 
C_{\alpha_s a|a_0 \cdots a_s}^* (\sigma_\alpha)^a_{~b} + \sum_{r = 0}^s
C_{\alpha_s b|a_0 \cdots a_{r - 1} a a_{r + 1} \cdots a_s}^* 
(\sigma_\alpha)^a_{~a_r},
\\
V_\alpha F_{\alpha_s|a_0 \cdots a_s} &= \sum_{r = 0}^s
F_{\alpha_s|a_0 \cdots a_{r - 1} a a_{r + 1} \cdots a_s} 
(\sigma_\alpha)^a_{~a_r}
\\
\intertext{and}
V \bar{A}_i &= - 2 \bar{A}_i,
\\
V A_{i b}^* &= -3 A_{i b}^*,
\\
V D_i &= - 4 D_i,
\\
V \bar{B}_{\alpha_s|a_1 \cdots a_s} &=
- (s + 4) \bar{B}_{\alpha_s|a_1 \cdots a_s},
\\
V B_{\alpha_s|a_1 \cdots a_s}^* &=
- (s + 5) B_{\alpha_s|a_1 \cdots a_s}^*,
\\
V E_{\alpha_s|a_0 \cdots a_s} &=
- (s + 6) E_{\alpha_s|a_0 \cdots a_s},
\\
V \bar{C}_{\alpha_s|a_0 \cdots a_s} &=
- (s + 3) \bar{C}_{\alpha_s|a_1 \cdots a_s},
\\
V C_{\alpha_s|a_0 \cdots a_s}^* &=
- (s + 4) C_{\alpha_s|a_0 \cdots a_s}^*,
\\
V F_{\alpha_s|a_0 \cdots a_s} &=
- (s + 5) F_{\alpha_s|a_0 \cdots a_s}.
\end{align*}
By an explicit calculation it can be verified that the generators
$V_\pm^a$, $V_\alpha$ and $V$ obey the $sl(1,2)$-superalgebra (\ref{II3}).

\end{appendix}


\end{document}